%% file: B2G-20-008_temp.tex
\pdfoutput=1
\documentclass[11pt,twoside,a4paper,cmspaper,final,collab]{cms-tdr}
\def\svnVersion{e5348a4}\def\svnDate{2022/07/14}\def\cmsCernNoTag{CERN-EP-2021-158}\def\cmsCernDate{\today}\def\cmsMessage{Published in Physical Review D as \href{http://dx.doi.org/10.1103/PhysRevD.106.012004}{\doi{10.1103/PhysRevD.106.012004}.}}
\begin{document}\cmsNoteHeader{B2G-20-008}

\newcommand{\mJ}{\ensuremath{m_\mathrm{J}}\xspace}
\DeclareRobustCommand{\PR}{{\HepParticle{R}{}{}}\xspace}
\newcommand{\WZtoqqbarprnunu}{\ensuremath{\PW\PZ \to \qqbar'\PGn\PAGn}\xspace}
\newcommand{\ZZtoqqbarnunu}{\ensuremath{\PZ\PZ \to \qqbar\PGn\PAGn\xspace}}
\ifthenelse{\boolean{cms@external}}{\providecommand{\cmsLeft}{upper\xspace}}{\providecommand{\cmsLeft}{left\xspace}}
\ifthenelse{\boolean{cms@external}}{\providecommand{\cmsRight}{lower\xspace}}{\providecommand{\cmsRight}{right\xspace}}

\cmsNoteHeader{B2G-20-008} 
\title{Search for heavy resonances decaying to \texorpdfstring{$\PZ(\PGn\PAGn)\PV(\PQq\PAQq')$}{Z(vvbar)V(qq')} in proton-proton collisions at \texorpdfstring{$\sqrt{s} = 13\TeV$}{sqrt(s) = 13 TeV}}

\author*[cern]{The CMS Collaboration}

\date{\today}

\abstract{
  A search is presented for heavy bosons decaying to \PZ{}(\PGn{}\PAGn)\PV{}($\PQq\PAQq'$), where \PV can be a \PW or a \PZ boson.  A sample of proton-proton collision data at $\sqrt{s}=13\TeV$ was collected by the CMS experiment during 2016--2018. The data correspond to an integrated luminosity of 137\fbinv.    The event categorization is based on the presence of high-momentum jets in the forward region to identify production through weak vector boson fusion.   Additional categorization uses jet substructure techniques and the presence of large missing transverse momentum to identify \PW and \PZ bosons decaying to quarks and neutrinos, respectively. The dominant standard model backgrounds are estimated using data taken from control regions. The results are interpreted in terms of radion, \PWpr boson, and graviton models, under the assumption that these bosons are produced  via gluon-gluon fusion, Drell--Yan, or weak vector boson fusion processes. No evidence is found for physics beyond the standard model.  Upper limits are set at 95\% confidence level on various types of hypothetical new bosons.  Observed (expected) exclusion limits on the masses of these bosons range from 1.2 to 4.0 (1.1 to 3.7)\TeV.  
}

\hypersetup{
pdfauthor={CMS Collaboration},
pdftitle={Search for heavy resonances decaying to Z(nu nubar)V(q qbar') in proton-proton collisions at sqrt(s) = 13 TeV},
pdfsubject={CMS},
pdfkeywords={CMS, diboson, extra dimensions, HVT, VBF, helicity}
}

\maketitle 

\section{Introduction}
\label{sec:intro}

The search for physics beyond the standard model (SM) using proton-proton ($\Pp\Pp$) collisions produced by the CERN LHC is 
a key goal of the CMS physics program. The apparent large hierarchy between the electroweak scale and the Planck scale, 
the nature of dark matter, and the possibility of a unification of the gauge couplings at high energies, 
are among the outstanding problems in particle physics not addressed within the SM. 
New bosons are predicted in many beyond-the-SM theories, which attempt to answer some of these questions.
New spin-0 and spin-2 particles are predicted in the Randall--Sundrum (RS) models of warped extra 
dimensions~\cite{Randall_1999a,Randall_1999b}, which arise from radion and graviton fields.
Spin-1 \PWpr and \PZpr bosons can also arise in these models~\cite{Agashe:2008jb,Agashe:2009bb,Agashe:2007ki},
as well as in left-right symmetric theories~\cite{Pappadopulo:2014qza}, and little Higgs models~\cite{ArkaniHamed:2002qx,ArkaniHamed:2002qy,Burdman:2002ns}. 
When the interaction strength between a new boson and the SM boson field is large, such as in the bulk RS models~\cite{Agashe:2007zd,Fitzpatrick:2007qr}, 
searches for diboson resonances are motivated, either via weak vector boson fusion (VBF) or strong production processes. 

\begin{figure*}[!htb]
        \centering
        \includegraphics[width=0.32\linewidth]{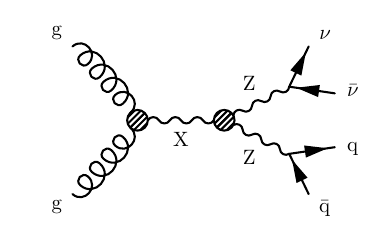}
        \includegraphics[width=0.32\linewidth]{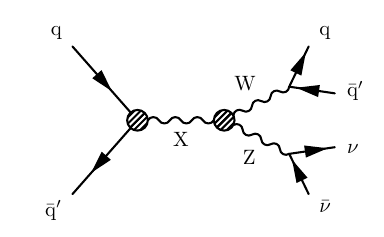}
        \includegraphics[width=0.32\linewidth]{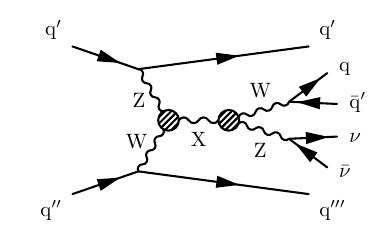}
        \caption{Representative Feynman diagrams for various production modes of a heavy resonance \PX. These modes are: a ggF-produced spin-0 or spin-2 resonance decaying 
    to {\ZZtoqqbarnunu} (left), a DY-produced spin-1 resonance decaying to \WZtoqqbarprnunu (center), and a VBF-produced spin-1 resonance decaying to \WZtoqqbarprnunu (right). }
        \label{fig:FeynDiag}
\end{figure*}

This paper presents a search for new bosons (\PX) decaying either to a pair of \PZ bosons or to a \PW and a \PZ boson, as shown in Fig.~\ref{fig:FeynDiag}.
In addition to VBF production, we consider gluon-gluon fusion (ggF) production of spin-0 and spin-2 particles, and Drell--Yan (DY) production of spin-1 particles.
The targeted final state is one in which one \PZ boson decays into a neutrino pair, while the other vector boson decays hadronically into jets.
Searches for similar signatures have been presented by the ATLAS~\cite{Aaboud:2017itg} and CMS~\cite{Sirunyan:2018ivv} Collaborations 
with the LHC data recorded in 2015 and 2016.  
Compared to earlier CMS publications, this paper includes additional data, new interpretations in terms of radion models, and extends 
the analysis to new search regions by including event categories that identify forward jets consistent with VBF production. 
An analysis similar to the one presented here was published by the ATLAS Collaboration using data recorded in 2015--2018~\cite{ATLAS:2020fry}.
In that analysis, no significant deviations from the SM were observed.

In the present work, proton-proton collisions at $\sqrt{s}=13\TeV$ recorded during 2016--2018 with the CMS detector are analyzed.  
The data set corresponds to an integrated luminosity of 137\fbinv.  
Events are selected that have a high-mass jet and a significant amount of missing transverse 
momentum (\ptmiss), which is the signature of the $\PZ\to\PGn{}\PAGn$ decay.
Events are categorized by the presence or absence of jets at large pseudorapidity (forward jets).  
Events that include forward jets are typical of VBF production.
Events without forward jets more commonly occur in ggF or DY processes. 
For sufficiently heavy new bosonic states, the decaying vector bosons will have high momenta and,
when decaying hadronically, will appear in the detector as single jets. 
As a result, we employ jet substructure techniques to identify these objects. 
Contributions from the dominant backgrounds are determined using control samples in data.

This paper is organized as follows.
Section~\ref{sec:detector} gives a brief description of the CMS detector and its components relevant to this work.
Section~\ref{sec:samples} summarizes the simulated data sets used in this analysis. 
Algorithms for the reconstruction of physics objects and the selection criteria applied to these objects are described in Sections~\ref{sec:reco} and~\ref{sec:selections}, respectively. 
Section~\ref{sec:bkg_est} discusses methods used for estimating the SM backgrounds.
The systematic uncertainties relevant to this analysis are described in Section~\ref{sec:syst}. 
The final results are presented in Section~\ref{sec:results}, along with their statistical interpretations. 
A summary of the work is presented in Section~\ref{sec:summary}. Tabulated results are provided in the HEPData record~\cite{hepdata} for this analysis. 

\section{The CMS detector}
\label{sec:detector}

The central feature of the CMS detector is a superconducting solenoid of 6\unit{m} internal diameter, providing a magnetic field of 3.8\unit{T}. 
Within the solenoid volume are a silicon pixel and strip tracker, a lead tungstate crystal electromagnetic calorimeter, and a brass and scintillator hadron calorimeter. 
Each of these systems is composed of a barrel and two endcap sections. 
The tracking detectors cover the pseudorapidity range $\abs{\eta}<2.5$. 
For the electromagnetic and hadronic calorimeters, the barrel and endcap detectors together cover the range $\abs{\eta}<3.0$. 
Forward calorimeters extend the coverage to $\abs{\eta}<5.0$. 
Muons are measured and identified in both barrel and endcap systems, which together cover the pseudorapidity range $\abs{\eta}<2.4$. 
The detection planes are based on three technologies: drift tubes, cathode strip chambers, 
and resistive-plate chambers, which are embedded in the steel flux-return yoke outside the solenoid. 
The detector is nearly hermetic, permitting accurate measurements of \ptmiss. 
Events of interest are selected using a two-tiered trigger system~\cite{Sirunyan:2020zal,Khachatryan:2016bia}. 
The first level, composed of custom hardware processors, uses information from the calorimeters and muon detectors to select events at a rate of around 100\unit{kHz} within a fixed time interval of less than 4\mus. 
The second level, known as the high-level trigger, consists of a farm of processors running a version of the full event reconstruction software optimized for fast processing, and 
reduces the event rate to around 1\unit{kHz} before data storage. 
A more detailed description of the CMS detector, together with a definition of the coordinate system used and the relevant kinematic variables, can be found in Ref.~\cite{Chatrchyan:2008zzk}.

\section{Simulated event samples}
\label{sec:samples}
This search uses Monte Carlo (MC) simulated data sets to study and guide the estimations of SM background processes.
Simulated data sets include: \PV{}+jets, where \PV refers to
\PW or \PZ bosons, \ttbar{}+jets, single top quark samples, diboson and 
triboson samples ($\PW\PZ$, $\PW\PZ\PZ$, etc.), and \ttbar{}+\PV samples.
Events in these samples for 2016 (2017 and 2018) are generated using \MGvATNLO v2.2.2 (v2.4.2)~\cite{Alwall:2014hca,Frederix:2012ps,Alwall:2007fs,Artoisenet:2012st}.
The \PV{}+jets (\ttbar{}+jets) samples are generated with leading order (LO) precision in the perturbative expansion of quantum chromodynamics (pQCD) 
and contain up to four (three) additional partons in the matrix element calculations.  
The $t$-channel and $\cPqt\PW$ single top quark events, and $\PW\PW$ events for 2016 (2017 and 2018) are 
generated using \POWHEG v1.0 (v2.0)~\cite{Nason:2004rx,Frixione:2007vw,Alioli:2010xd,Alioli:2009je,Re:2010bp} at next-to-LO (NLO) precision in pQCD. 
All other SM background samples are generated using \MGvATNLO at NLO precision in pQCD.
The signal samples used for interpretation of the results are also generated using \MGvATNLO with LO precision in pQCD.

The parton showering and hadronization step in all simulations is performed with two versions of 
\PYTHIA~\cite{Sjostrand:2014zea}. The 2016 (2017 and 2018) simulation uses the 
CUETP8M1~\cite{Khachatryan:2015pea} (CP5~\cite{Sirunyan:2019dfx}) underlying event tune with PYTHIA v8.212 (v8.230).
The parton distribution functions (PDFs) used in the 2016 (2017 and 2018) simulated data samples are 
NNPDF3.0 LO or NLO~\cite{Ball:2014uwa} (NNPDF3.1 NNLO~\cite{Ball:2017nwa}). 
The simulation of the particle interactions and the CMS detector is performed with \GEANTfour~\cite{Agostinelli:2002hh}. 
The effects due to additional $\Pp\Pp$ interactions in the same or adjacent bunch crossings (pileup) are also simulated and the events are
weighted to match the pileup distribution in data. 

The cross sections used to normalize different SM processes correspond to 
NLO or next-to-NLO (NNLO) accuracy~\cite{Alioli:2009je,Re:2010bp,Alwall:2014hca,Melia:2011tj,Beneke:2011mq,Cacciari:2011hy,Baernreuther:2012ws,Czakon:2012zr,Czakon:2012pz,Czakon:2013goa,Gavin:2012sy,Gavin:2010az}.
The \PV{}+jets samples are weighted based on the transverse momentum (\pt) of the \PW and \PZ bosons.  Weights are derived from comparisons of LO simulations and
simulations at NNLO precision in pQCD interactions and NLO precision in electroweak interactions~\cite{Lindert:2017olm}.  

\begin{figure}[!htbp]
  \centering
    \includegraphics[width=0.45\textwidth]{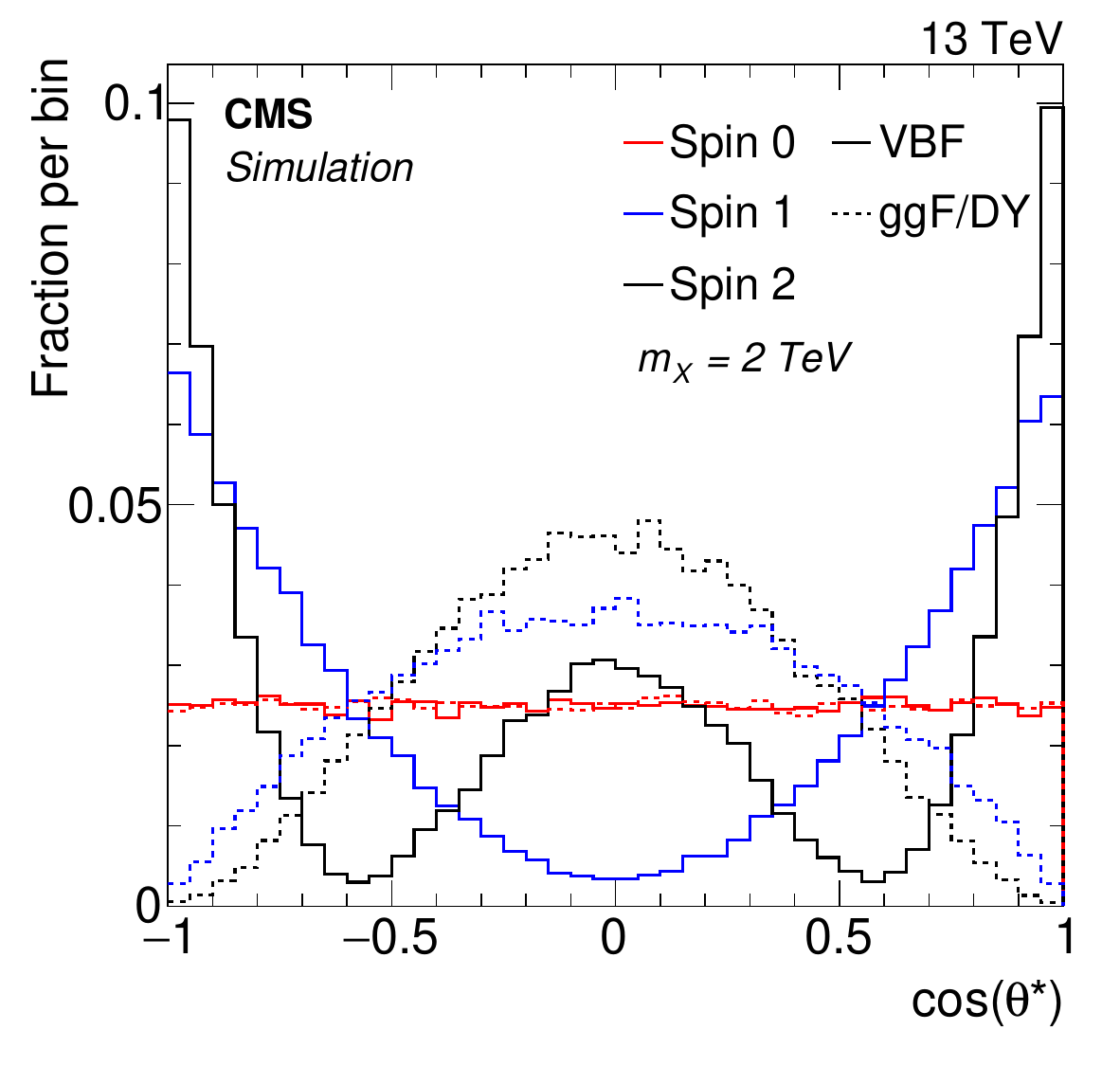}
    \caption{Simulated distributions are shown for the cosine of the decay angle of SM vector bosons in the rest frame of 
      a parent particle with a mass ($m_\PX$) of 2\TeV. Solid lines represent VBF scenarios. Dashed lines represent ggF/DY scenarios. The integral of
    each histogram is normalized to unity.}
  \label{fig:cos_theta_star}
\end{figure}

When diboson signatures are produced through VBF, correlations between the spin of initial- and final-state
bosons cause event kinematic distributions to depend on the hypothesized resonance spin.
These correlations are manifested in the decay angle, $\theta^{*}$, of the vector bosons, defined
in Ref.~\cite{Bolognesi:2012mm}.  Figure~\ref{fig:cos_theta_star}
shows the distribution of the cosine of the decay angles in the rest frame of \PX for each spin configuration 
and production mechanism we consider, given a parent particle mass of 2\TeV.
Scalar resonances have no correlation between initial- and final-state spins, which translates to a flat $\cos\theta^{*}$ distribution.  
Spin-1 and spin-2 particles can have strong correlations, especially when produced via VBF, where vector bosons are preferentially produced in the forward direction ($\cos\theta^{*}\simeq\pm 1$).
Such correlations are caused by the production of longitudinally or transversely polarized vector bosons~\cite{Oliveira:2014kla}.
The decay angle is correlated with the \pt of the resonance's decay products.
Thus, the spectral features explored here are different from typical diboson searches because the final state is only partially reconstructed.
We explore the sensitivity of the analysis to the assumed spin by considering three different production models: radions (spin 0), \PWpr bosons (spin 1), and gravitons (spin 2).  
In each case, we consider both VBF and ggF or DY production, and masses between 1.0 and 4.5\TeV in steps of 200 (500)\GeV for masses less (greater) than 2.0\TeV.

To interpret the data in terms of spin-0 and spin-2 signals, radions and gravitons arising from models of warped extra dimensions~\cite{Randall_1999a,Randall_1999b},
and specifically from bulk scenarios~\cite{Agashe:2007zd,Fitzpatrick:2007qr},
are used as representative models. The warped extra dimension model has several free parameters,
including the mass of the lowest excited state of the graviton, the radion mass, 
the parameter $kr_{c}\pi$, where $k$ and $r_{c}$ are the curvature and the compactification scale of the extra dimension respectively, 
the dimensionless coupling $\widetilde{k}=k/\overline{M_{\mathrm{Pl}}}$, where $\overline{M_{\mathrm{Pl}}}$ is the reduced Planck mass,
and the energy scale associated with radion interactions ($\Lambda_{\mathrm{R}}$).
These parameters and the corresponding cross sections are described in Ref.~\cite{Oliveira:2014kla}.
We assume $\widetilde{k}=0.5$, $kr_{c}\pi=35$, and $\Lambda_{\mathrm{R}}=3\TeV$, which
results in decay widths for gravitons and radions smaller than 10\% of their mass.  
The predicted cross sections for radions and gravitons produced via ggF (VBF) are accurate to NLO (LO) in pQCD.  

To interpret data in terms of  spin-1 signals, \PWpr bosons within the heavy vector triplet (HVT) framework~\cite{Pappadopulo:2014qza}
are used.
The interactions between
\PWpr bosons and SM particles are parameterized in terms of $c_{\mathrm{H}}$,
$g_{\mathrm{V}}$, and $c_{\mathrm{F}}$ couplings and the mass of the HVT \PWpr boson.  Together with the SM $\mathrm{SU}(2)_{\mathrm{L}}$ gauge coupling, $g$,
these parameters determine the couplings between the HVT bosons and
the SM Higgs boson, SM vector bosons, and SM fermions, respectively, according to Ref.~\cite{Pappadopulo:2014qza}.
We consider two specific cases of these model parameters: those corresponding to
model B ($c_{\mathrm{H}}=-0.98$, $g_{\mathrm{V}}=3$, and $c_{\mathrm{F}}=1.02$) in Ref.~\cite{Pappadopulo:2014qza} and model C ($c_{\mathrm{H}}=g_{\mathrm{V}}=1$ and $c_{\mathrm{F}}=0$).
Model B suppresses the fermion couplings, enhancing decays to
SM vector bosons. Model C removes direct couplings to fermions, ensuring that only VBF production is possible. 
The predicted cross sections for \PWpr bosons are accurate to LO in pQCD.

\section{Triggers and event reconstruction}
\label{sec:reco}

Samples of collision data are selected using several \ptmiss triggers whose thresholds vary between 90 and 140\GeV, depending on the data-taking period.
The \ptmiss trigger efficiencies were studied using samples of events selected with single-lepton triggers. 
The lepton triggers isolate events with neutrinos produced via \PW boson decays, which serve as a good proxy for both the signal and the dominant backgrounds.  
The single-lepton triggers required electrons with $\pt > 27\text{--}32\GeV$, depending on the data-taking period, or muons with $\pt > 24\GeV$. 
The efficiencies of the triggers for selecting events with large \ptmiss ($>$200\GeV) were measured as functions of \ptmiss.  
The differences between the efficiencies determined using electron events and muon events were found to be less than 1.4\% 
and are assigned as systematic uncertainties.
For all data-taking periods, the trigger efficiency is found to reach 95\% for $\ptmiss > 250\GeV$ and plateaus around 98\% for $\ptmiss > 300\GeV$. 
At $\ptmiss=200\GeV$, the efficiency is found to be greater than 75\%. 
The trigger efficiencies measured using single-lepton data are applied to the simulated data as functions of the reconstructed \ptmiss.
The dependence of these efficiencies on other kinematic variables has been investigated and the associated effects found to be less than the assigned uncertainties.

The event reconstruction proceeds from particles identified by the particle-flow (PF) algorithm~\cite{CMS-PRF-14-001}, which uses 
information from the silicon tracking system, calorimeters, and muon systems to reconstruct PF candidates as electrons, muons, charged or neutral hadrons, or photons. 
The candidate vertex with the largest value of summed physics-object $\pt^2$ is taken to be the primary $\Pp\Pp$ interaction vertex. 
The physics objects used for this determination are jets, formed by clustering tracks assigned to candidate vertices using the anti-\kt jet 
finding algorithm~\cite{Cacciari:2008gp,Cacciari:2011ma} with a distance parameter of 0.4, and the associated \ptvecmiss, 
taken as the negative vector \pt sum of those jets.

Electrons are reconstructed by associating a charged-particle track with an electromagnetic calorimeter supercluster~\cite{Khachatryan:2015hwa}. 
The resulting electron candidates are required to have $\pt>10\GeV$ and $\abs{\eta}<2.5$, and to satisfy identification criteria designed to reject light-parton jets, photon conversions, and electrons produced in the decays of heavy-flavor hadrons. 
Muons are reconstructed by associating tracks in the muon system with those found in the silicon tracker~\cite{Sirunyan:2018fpa}. 
Muon candidates are required to satisfy $\pt>10\GeV$ and $\abs{\eta}<2.4$.

Leptons (\Pe,~\PGm) are required to be isolated from other PF candidates to select preferentially those originating from \PW and \PZ boson decays and suppress backgrounds in which leptons are produced in the decays of hadrons containing heavy quarks.
Isolation is quantified using an optimized version of the mini-isolation variable originally introduced in Ref.~\cite{Rehermann:2010vq}. 
The isolation variable, $I_{\text{mini}}$, is calculated by summing the $\pt$ of the charged and neutral hadrons, and photons with 
$\Delta R\equiv\sqrt{\smash[b]{(\Delta\phi)^2+(\Delta\eta)^2}}<R_0$ of the lepton momentum vector, $\vec{p}^{\kern1.5pt\Pell}$, where $\phi$ is the azimuthal angle and $R_0$ depends on the lepton \pt.
The three values of $R_0$ used are: 0.2 for $\pt^{\Pell}\leq 50\GeV$; $(10\GeV)/\pt^{\Pell}$ for $50<\pt^{\Pell}<200\GeV$; and 0.05 for $\pt^{\Pell}\geq 200\GeV$. 
Electrons (muons) are then required to satisfy $I_{\text{mini}}/\pt^{\Pell}<0.1\,(0.2)$. 
In order to mitigate the effects from pileup in an event, charged hadron candidates are required to originate 
from the primary vertex of the given event.

Events are vetoed in which muon or electron candidates are identified that satisfy isolated track requirements, $\abs{\eta}<2.4$, and $\pt>5\GeV$, 
but do not satisfy the kinematic requirements just described for isolated leptons.
Hadronically decaying \PGt leptons can produce isolated hadron tracks.
Events in which an isolated hadron track is identified that satisfies $\abs{\eta}<2.4$ and $\pt>10\GeV$ are also vetoed.

Events with one or more photons are vetoed. Photon candidates are required to have $\pt>100\GeV$, $\abs{\eta}<2.4$, and be isolated from neutral
hadrons, charged hadrons, and electromagnetic particles, excluding the photon candidate itself.
The isolation criterion is calculated using a cone with $\Delta R<0.2$ around the photon~\cite{Khachatryan:2015iwa}.  

Jets are reconstructed by clustering charged and neutral PF candidates using the anti-\kt algorithm.  
Two collections of jets are considered clustered with distance parameters $R=0.4$ and 0.8, as implemented in the \FASTJET package~\cite{Cacciari:2011ma}. 
Depending on the distance parameter used, these jets are referred to as AK4 and AK8 jets, respectively. 
Jet energies are corrected using a \pt- and $\eta$-dependent jet energy calibration~\cite{Khachatryan:2016kdb}.
To mitigate contributions to jet reconstruction from pileup, we utilize
the charged-hadron subtraction method for AK4 jets and the pileup-per-particle 
identification (PUPPI) method~\cite{Bertolini:2014bba,Sirunyan:2020foa} for AK8 jets.

Jets are required to satisfy $\pt>30\,(200)\GeV$, $\abs{\eta}<5.0\,(2.4)$, and jet quality criteria~\cite{cms-pas-jme-10-003,CMS-PAS-JME-16-003} for AK4 (AK8) jets. 
The mass of AK8 jets and information about the pattern of energy deposited in various detectors are used to reconstruct massive vector bosons.  
To further reduce the dependence on pileup and to help reduce the effect of wide-angle soft radiation, the soft-drop algorithm~\cite{Dasgupta:2013ihk,Larkoski:2014wba} is used to remove constituents from the AK8 jets. 
Corrections are applied to AK8 jets to ensure that the reconstructed jet mass reproduces the pole masses of the SM bosons.
Corrections are derived using \PW bosons from \ttbar production to account for known differences between measured and simulated jet mass scale and jet mass resolution~\cite{Sirunyan_2020}.
Finally, jets having PF constituents matched to an isolated lepton are removed from the jet collection.

The $N$-subjettiness~\cite{Thaler:2010tr} technique is used to distinguish between AK8 jets originating from the hadronic decay of \PW or \PZ bosons and those originating from quantum chromodynamics (QCD).
In hadronic boson decays, the resulting jet is likely to have two substructure components, which results in a smaller $N$-subjettiness ratio, $\tau_{21}$. 
For QCD jets, $\tau_{21}$ tends to be higher. We require AK8 jets to satisfy $\tau_{21}<0.75$. 
There is a loss of about 36\% in background events due to this requirement. Negligible loss of efficiency is observed due to 
this requirement in a signal sample of VBF-produced gravitons (VBFG) with a mass of 1\TeV. 
The efficiency for tagging \PW bosons in simulated data sets is validated against collision data using \ttbar events~\cite{Sirunyan_2020}.  
Scale factors are applied to simulated data sets to account for observed differences. 

The AK4 jets are tagged as originating from the hadronization of \PQb quarks using the deep combined secondary vertex 
algorithm~\cite{Sirunyan:2017ezt}. 
For the medium working point chosen here, the signal efficiency for identifying \PQb jets with $\pt>30\GeV$ in $\ttbar$ events is about 68\%. 
The probability of misidentifying jets in $\ttbar$ events arising from \PQc quarks is approximately 12\%, 
while the probability of misidentifying jets associated with light-flavor quarks or gluons as \PQb jets is approximately 1\%. 

The vector \ptvecmiss is defined as the negative vector \pt sum of all PF candidates and 
is calibrated taking into account the jet energy 
corrections. Its magnitude is denoted by \ptmiss. 
Dedicated event filters designed to reject instrumental noise are applied to improve the correspondence between
the reconstructed and the genuine \ptmiss~\cite{Sirunyan:2019kia}.
To suppress SM backgrounds,
we use the quantity \mT, defined as the transverse mass
of the system consisting of the highest \pt AK8 jet ($\pt^{\mathrm{J}}$) and \ptvecmiss.  The value of \mT is computed as
\begin{linenomath}
\begin{equation} 
\mT = \sqrt{{2\pt^{\mathrm{J}}\ptmiss[1-\cos\Delta\Phi]}}, 
\label{eq:MT}
\end{equation}
\end{linenomath}
where $\Delta\Phi$ is the difference between the azimuthal angle of the AK8 jet's momentum 
and \ptvecmiss. The \mT variable in Eq.~\ref{eq:MT} assumes that the 
new state decays into much lighter daughter particles.
 
\section{Event selection}
\label{sec:selections}

The reconstructed objects (defined in Section~\ref{sec:reco}) are used to define signal regions (SRs) and control regions (CRs). 
These are all subsets of a baseline phase space where at least one AK8 jet and 
$\ptmiss>200\GeV$ are required.  Events are rejected if they contain a reconstructed electron, muon, photon,
isolated track, or an AK4 jet which has been identified as a \PQb jet.  
Events are vetoed if any jet passes the \pt and $\eta$ requirements, but fails the jet quality criteria.
Events are also vetoed if \ptvecmiss is aligned with the transverse projection of any of the four highest \pt AK4 jets, defined as $\Delta\phi(j_i,\ptmiss)<0.5$.

The baseline phase space is split into two sets of regions, VBF and ggF/DY.  
The VBF regions require at least two AK4 jets, each required to be separated by $\Delta R>0.8$ from the AK8 jet. 
The two highest-\pt AK4 jets are also required to be reconstructed in opposite hemispheres ($\eta_1\eta_2<0$), 
have a large pseudorapidity separation ($\Delta\eta>4.0$), and form a large invariant mass ($m_{jj}>500\GeV$).
These requirements are indicative of VBF production. If any of these conditions is failed, the event is put into the ggF/DY category. 

Both the VBF and ggF/DY categories are further separated into
high-purity (HP) and low-purity (LP) categories, depending on the
highest \pt AK8 jet's value of $\tau_{21}$.
Events in which the highest \pt AK8 jet satisfies $\tau_{21}<0.35$ ($0.35<\tau_{21}<0.75$)
are referred to as HP (LP). 
The signal (VBFG 1\TeV) selection efficiency with the high (low) purity cut is 69 (31)\%. 
The background selection efficiency with the high (low) purity cut is 32 (68)\%.

Each of these four event categories is divided into a number of \mT bins.
Starting from $\mT=400\GeV$, each bin has a width of 100\GeV. In the ggF/DY (VBF) HP
category, the bin width is constant up to 2200 (1600)\GeV, beyond which, bin boundaries correspond
to $\mT=$ 2200, 2350, 2550, 2750, and 3000 (1600, 1750, and 2075)\GeV.  In the ggF/DY (VBF) LP
category, the bin width is constant up to 2900 (2300)\GeV, beyond which bin widths are 200\GeV up to
$\mT=3500$ (2700)\GeV.  The last bin in each of the categories includes all events with \mT above 
the final quoted bin boundary.  

In all \mT bins, an SR and a CR is defined based on the mass of the highest \pt AK8 jet, \mJ.  
The SRs require $65<\mJ<105\GeV$, a range which is chosen to accept both \PW and \PZ bosons, 
but reject most hadronically decaying Higgs bosons.  The CRs require either
$30<\mJ<65\GeV$ or $135<\mJ<300\GeV$, which excludes the SR mass requirement and a window around 
the Higgs boson mass. The event selections used in the analysis are summarized in Table~\ref{tab:selection}.
\begin{table*}[!htbp]
    \centering
    \topcaption{Summary of the event selections.}
    \begin{scotch}{ll}
    Variable                                        &           Selection                                               \\
    \hline
    \ptmiss                                         &  $>$200\GeV                                                    \\
    Veto                                            &  electrons, muons, tau leptons, photons, \PQb jets                   \\        
    AK4 jet \pt                                     &  $>$30\GeV                                                       \\
    $\Delta\phi(\ptvec^{\text{jets}},\ptvecmiss)$   &  $>$0.5                                                         \\
    Leading AK8 jet \pt and $\eta$                  &  $>$200\GeV and $\abs{\eta} < 2.4$                               \\
    Leading AK8 jet mass                            & SR: 65--105\GeV, CR: 30--65\GeV or 135--300\GeV  \\
    $\tau_{21}$                                     & HP: $<$0.35, LP: 0.35--0.75                                      \\             
    Forward jets                                    & $(\eta_1\eta_2)<0$, $\Delta\eta>4.0$, $m_{\text{jj}}>500\GeV$     \\
    \end{scotch}
    \label{tab:selection}
\end{table*}

The dominant backgrounds are events originating from \PW{}+jets and \PZ{}+jets production, followed by the 
$\PW\to\Pell\PGn$ and $\PZ\to\PGn\PAGn$ decays.
In these events the charged leptons do not pass the reconstruction requirements described
in Section~\ref{sec:reco}, while the neutrinos manifest themselves as \ptmiss.
In addition, the massive jets in these events do not arise from hadronically-decaying vector bosons;
rather they arise from the tail of a smoothly falling distribution of jet mass. We refer to these and other
processes with similar kinematic properties as nonresonant backgrounds.  The shape and normalization of nonresonant
backgrounds are constrained using data, as described in
Section~\ref{sec:bkg_est}.

Subdominant SM contributions come from \ttbar, single top quark, and
diboson processes. These processes typically have a hadronically-decaying
vector boson. There is a small contribution from the diboson events produced via the vector boson scattering process.  
These events are expected to contribute less than 10\% of the total diboson event
yield in the VBF SRs.  The contributions from subdominant backgrounds, referred to as resonant backgrounds, are estimated using
predictions from simulation.

\begin{figure}[!htbp]
  \centering
  \includegraphics[width=0.45\textwidth]{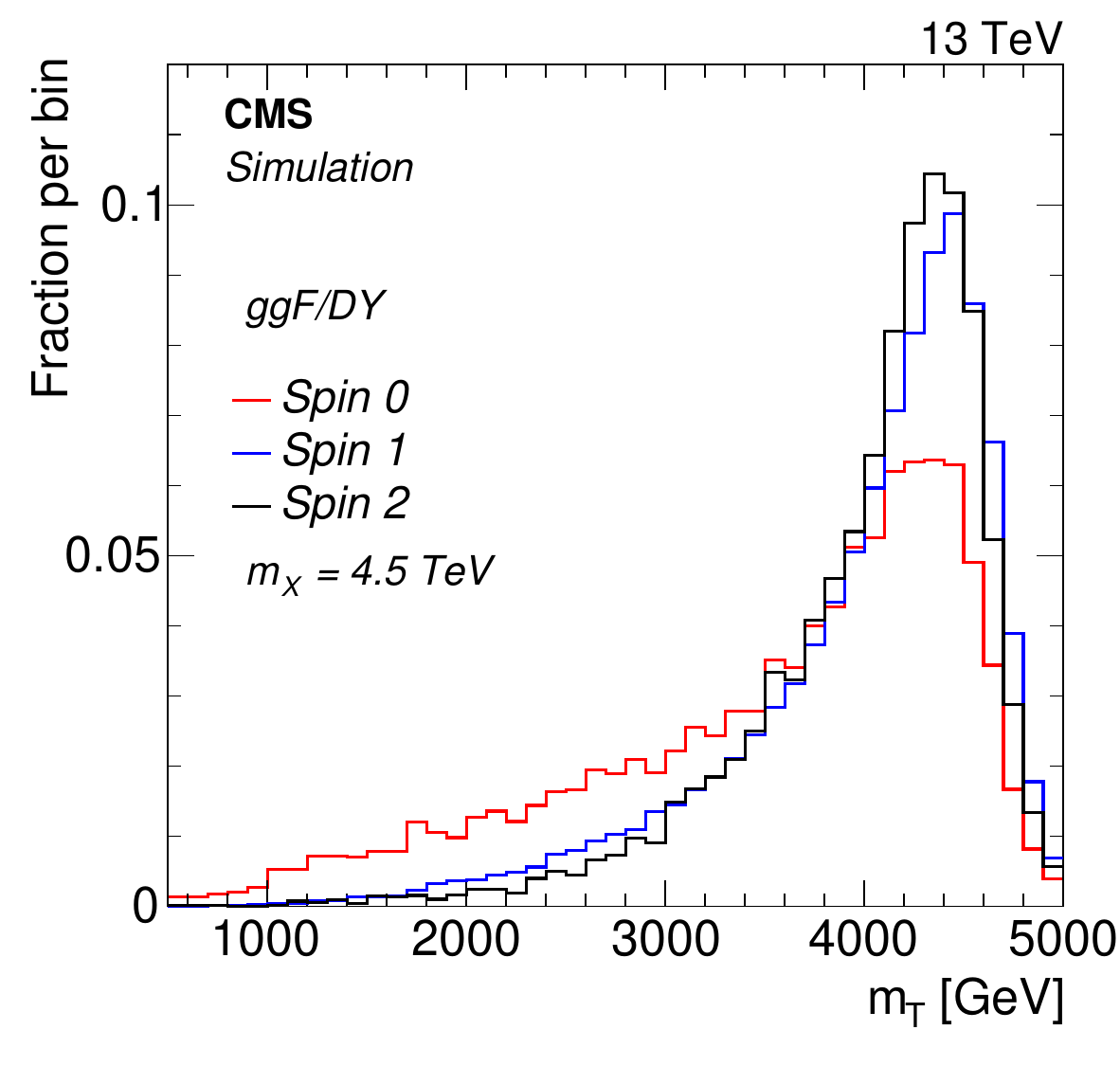}
  \includegraphics[width=0.45\textwidth]{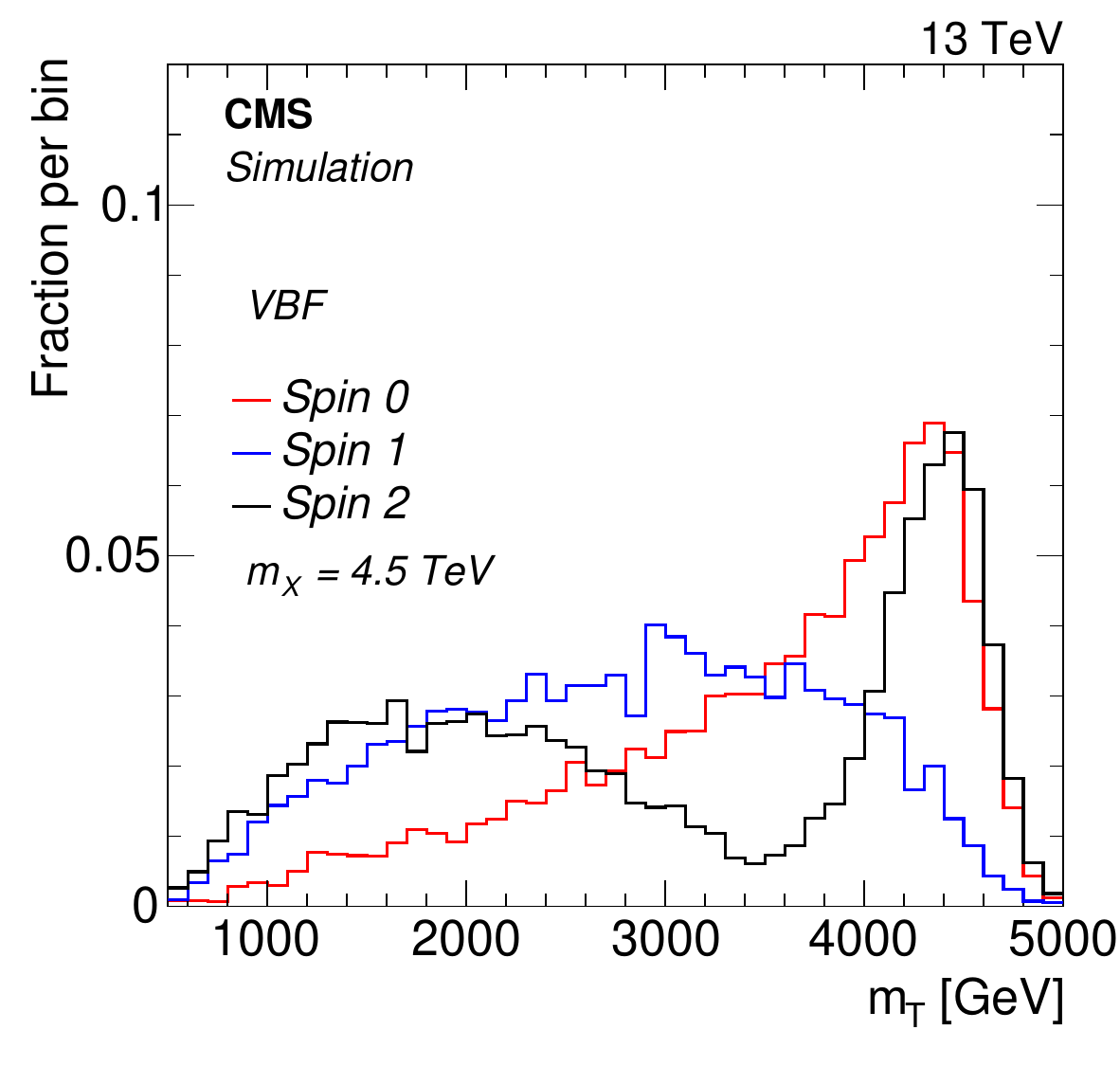}
  \caption{Distributions of \mT for ggF/DY- (\cmsLeft) and VBF-produced (\cmsRight) resonances \PX of mass 4.5\TeV.
            Events used are from all SR and CR combined. The integral of each histogram is normalized to unity.}
  \label{fig:signal_mt}
\end{figure}

The reconstructed \mT distribution for the signal depends on the spin, mass, and production mechanism. 
Using events from all signal 
regions and control regions combined, Fig.~\ref{fig:signal_mt} shows the \mT distributions for the
various signal hypotheses considered.  These distributions for VBF-produced \PWpr and graviton resonances
are a reflection of the prevalence for vector bosons production at large $\eta$ due to correlations between
$\eta$ and $\cos\theta^{*}$.  When vector bosons have large $\eta$, the AK8 jet \pt and the \ptmiss are lower,
and in turn, the reconstructed \mT is lower.
For models with multimodal $\cos{}\theta^{*}$ distributions, there is a corresponding
distortion in the \mT distribution, which tends to produce higher yields at
lower values of \mT.

\section{Background estimation method}
\label{sec:bkg_est}

\begin{figure}[!htbp]
  \centering
  \includegraphics[width=0.45\textwidth]{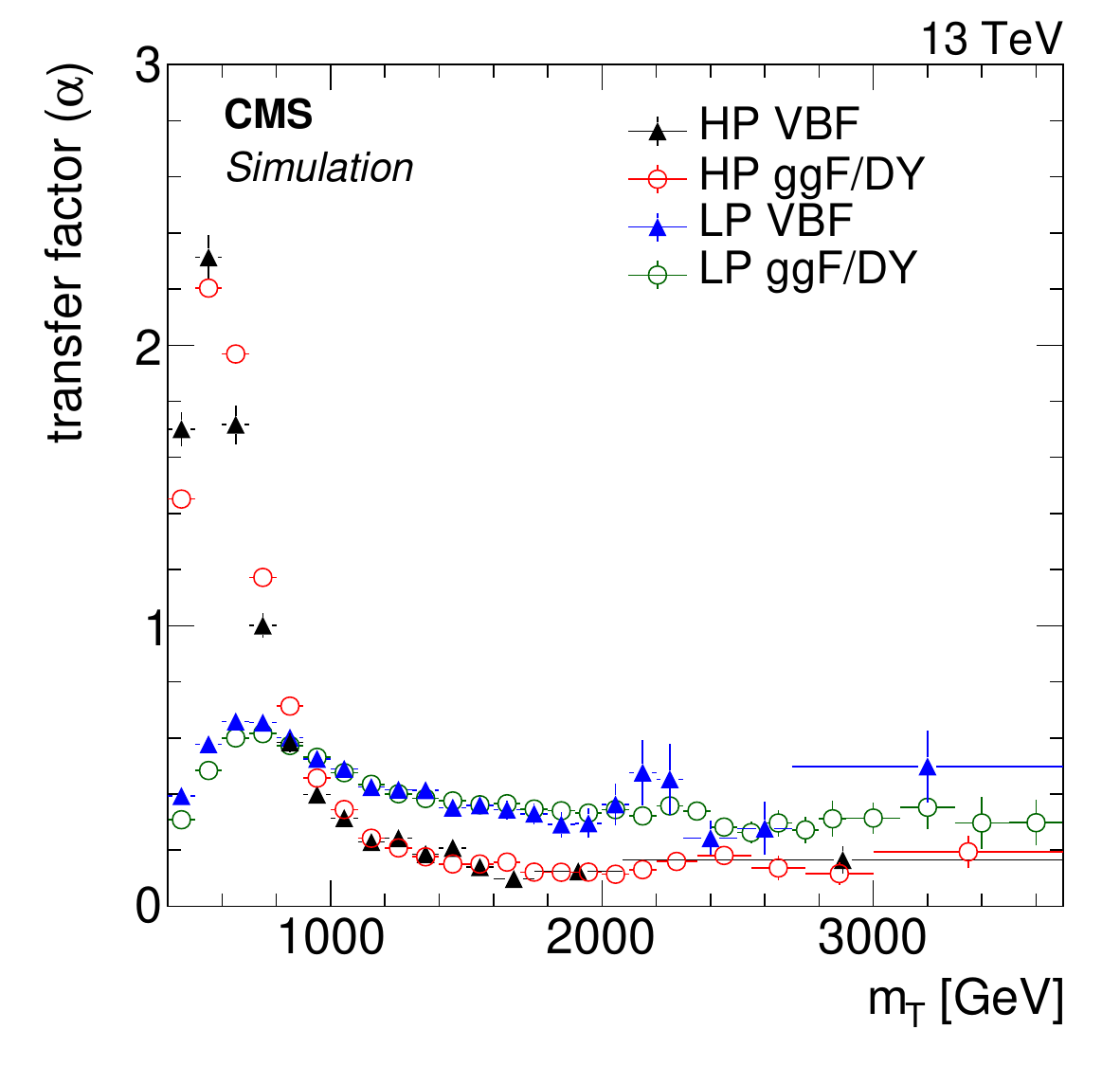}
  \caption{The distributions of the transfer factors ($\alpha$) versus \mT in the various event categories are shown. 
    The last bin corresponds to the value obtained by integrating events above the penultimate bin.}
  \label{fig:Alpha_Run2}
\end{figure}

The nonresonant backgrounds populate both the CRs and the SRs, while signal events mainly populate the SRs.
The yields observed in the CRs are weighted by a transfer factor, $\alpha$, to account for known differences between the SR and CR kinematic properties.
The transfer factor is derived from simulated data sets.  Algebraically, the predictions of the nonresonant backgrounds are given by
\begin{equation}
\begin{gathered}
    N_{\text{pred}}^{\text{non-res}} = \alpha(N_{\text{CR}}^{\text{obs}}-N_{\text{CR}}^{\text{res}}), \\
    \alpha=\frac{N_{\text{SR}}^{\text{non-res}}}{N_{\text{CR}}^{\text{non-res}}},
\end{gathered}
\label{eq:alpha}
\end{equation}
where $N_{\text{CR}}^{\text{obs}}$ refers to the observed CR yields, and 
$N^{\text{non-res}}$ ($N^{\text{res}}$) 
refers to the nonresonant (resonant) background yields in simulated data sets.
In this formula, $N_{\text{*}}^{\text{res}}$ serves the role of removing the expected contributions from resonant backgrounds, which have a systematically different transfer factor and whose predictions are handled separately.
All event yields are derived separately for each \mT bin and each event category.
Figure~\ref{fig:Alpha_Run2} shows $\alpha$ as a function of \mT in each of the event categories.

The resonant backgrounds are directly estimated using predictions from simulation, with systematic uncertainties evaluated to account for potential data mismodeling.
The resonant background yields are used to account for contamination in the CRs before predicting the nonresonant backgrounds and their
contribution to the SRs themselves.

\begin{figure*}[!htbp]
  \centering
    \includegraphics[width=0.45\textwidth]{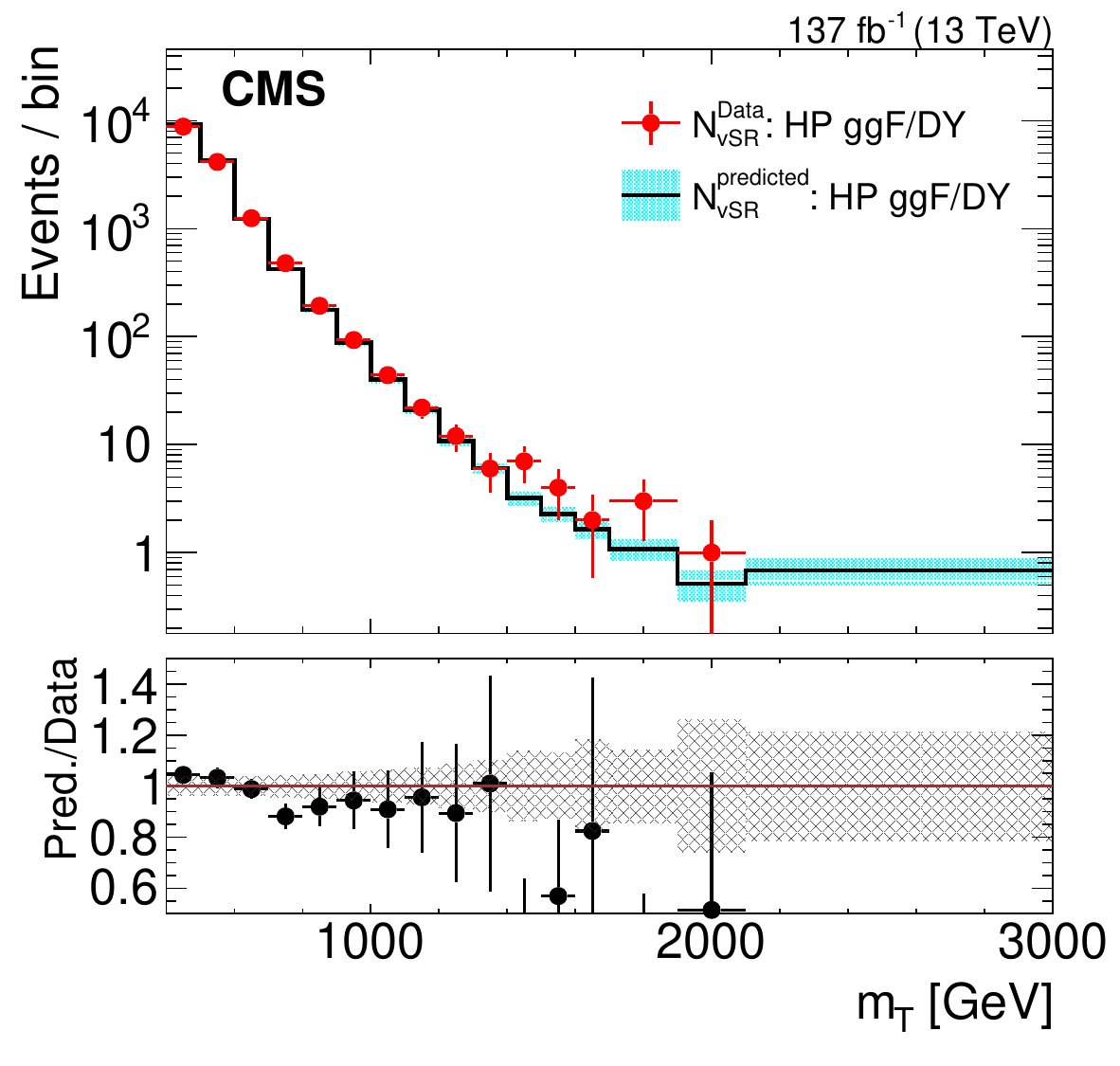}
    \includegraphics[width=0.45\textwidth]{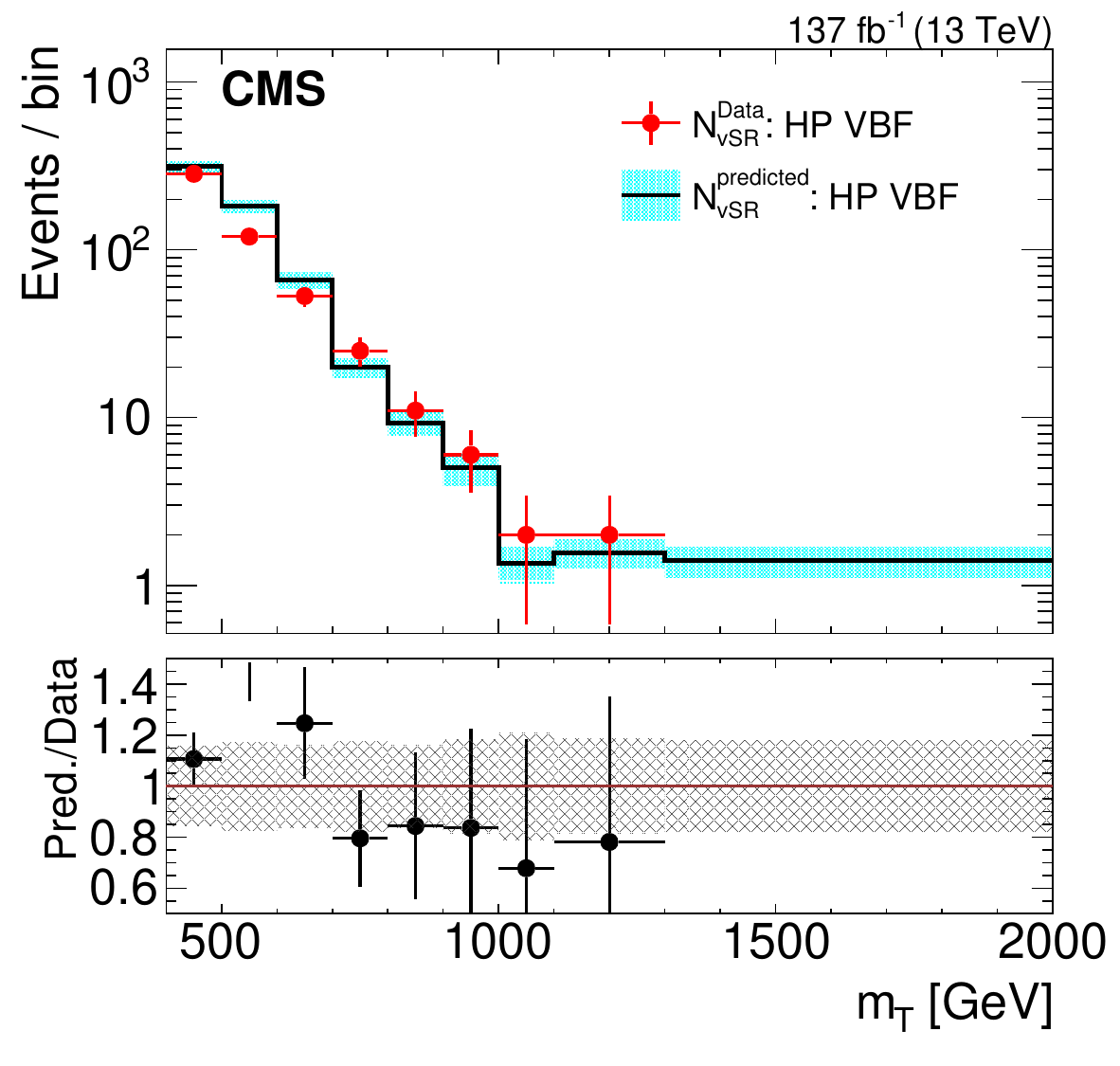}\\
    \includegraphics[width=0.45\textwidth]{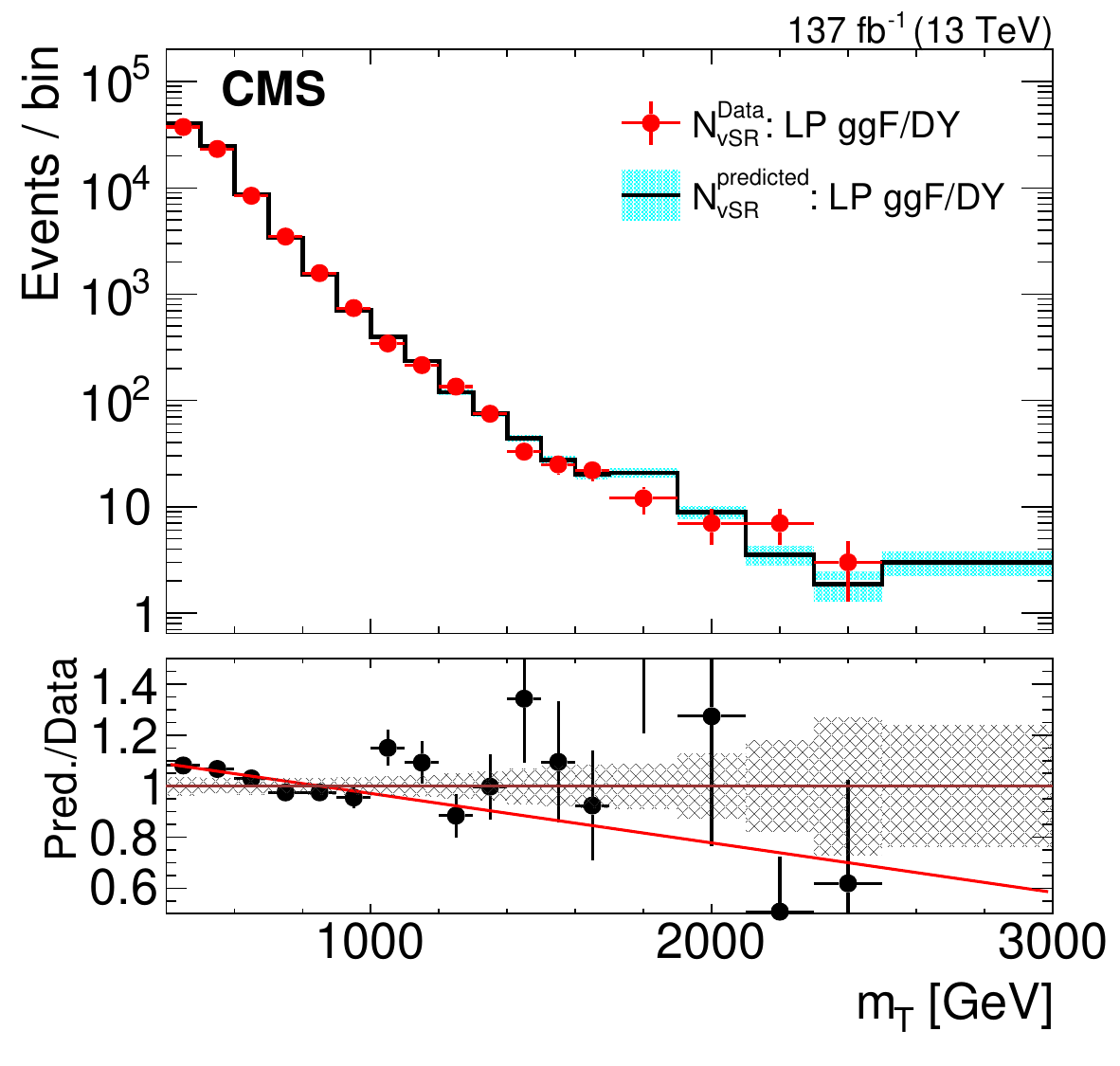}
    \includegraphics[width=0.45\textwidth]{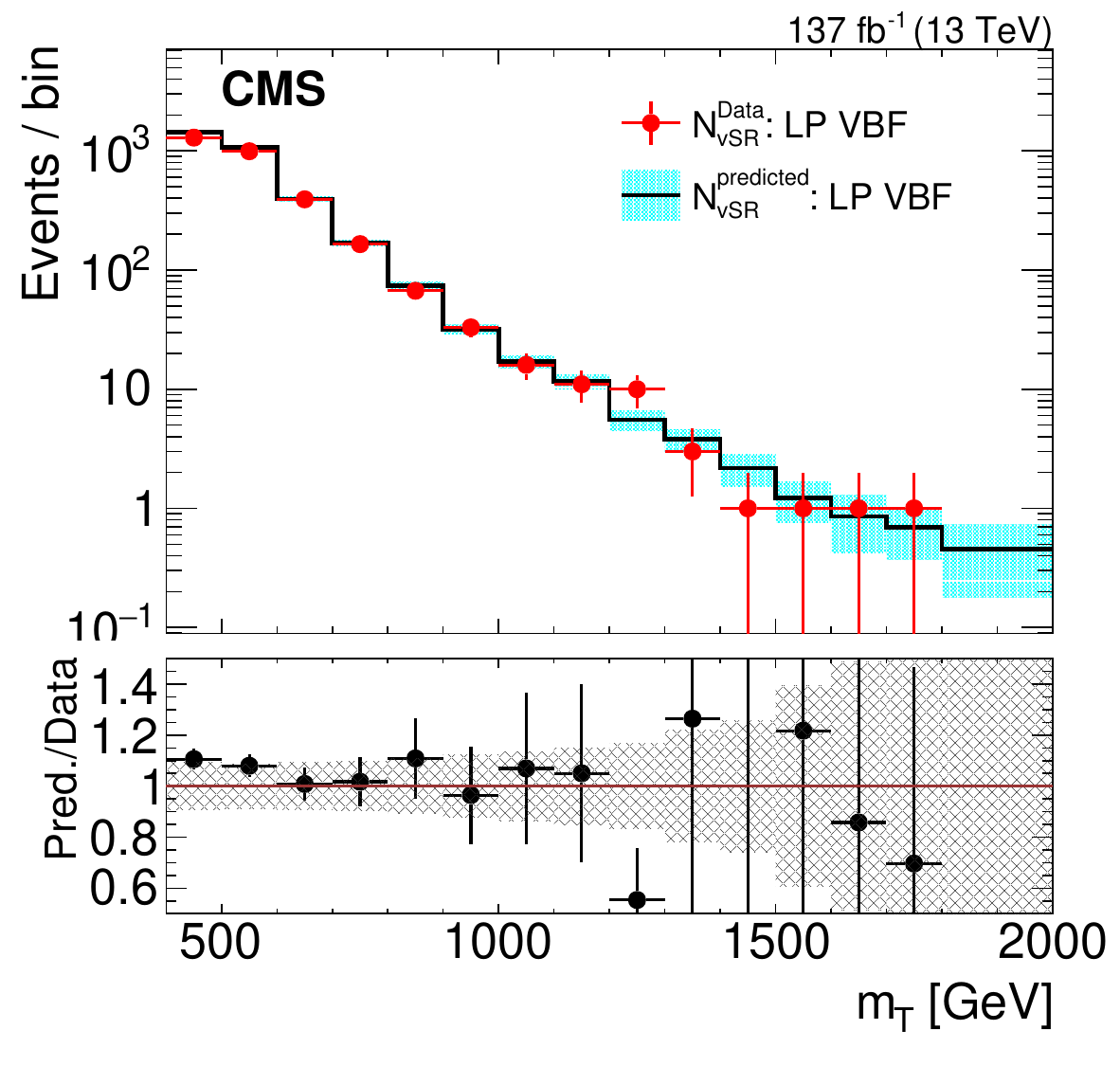}
    \caption{Comparison of background estimations and observations in the high-purity ggF/DY (upper left), high-purity VBF (upper right), low-purity ggF/DY (lower left), and low-purity VBF (lower right) validation signal regions.
      The lower panel shows the ratio of the estimated and the observed event yields.
      The hashed band in the ratio represents the total uncertainty in the corresponding SR. 
      The red line (lower left) is a fit to the ratio of prediction to the data in the LP ggF/DY vSR.}
  \label{fig:validation}
\end{figure*}

The procedures used to predict SM backgrounds have been validated using data
from a subset of each CR, which we refer to as the validation SR (vSR). The vSRs have the same
selections as the SRs except that the jet mass must satisfy $55<\mJ<65\GeV$. 
The CRs for validation tests are redefined by removing the events in the vSR.
The vSR and analogous CRs are used to compute the analog of $\alpha$, and the full background prediction is evaluated on data.
The predicted yields in the vSR are then compared to the observed event yields. This comparison is performed in the corresponding vSR of each of the event categories. 
The resulting distributions are shown in Fig.~\ref{fig:validation}.

The vSR tests result in a prediction of more events in each of the event categories than observed in the data. The overprediction is as large as 10\% in the lowest \mT bin.
To account for any potential mismodeling of our prediction, we derive a shape uncertainty.  The uncertainty
is based on a linear fit to the ratio of the prediction and the observation versus \mT in the LP ggF/DY vSR.
Based on the fit, an uncertainty is assessed that corresponds to 7\%  in the lowest \mT bin, and decreases linearly to -40\% in the highest \mT bin.
This shape uncertainty is applied to each of the four signal regions assuming no correlation between the regions.  

\section{Systematic uncertainties}
\label{sec:syst}

Because the nonresonant backgrounds are estimated from data, several simulation-related
systematic uncertainties have little to no effect on the estimation of these backgrounds.  
However, PDF uncertainties, renormalization ($\mu_\mathrm{R}$) and factorization ($\mu_\mathrm{F}$) scale uncertainties, 
and jet energy scale (JES) and resolution (JER) uncertainties have nonnegligible effects on $\alpha$.

For PDF uncertainties, the distribution of $\alpha$ is evaluated for each recommended PDF variation~\cite{Ball:2014uwa, Ball:2017nwa}.
An envelope of these variations is used to assess an $\alpha$ shape uncertainty.
The size of the uncertainty in a given \mT bin due to PDFs is as large as 3 (1.5)\% for the VBF (ggF/DY) categories.
The effects of the scale uncertainties are evaluated by varying the 
values of $\mu_\mathrm{R}$ and $\mu_\mathrm{F}$ simultaneously up and down by a factor of two.
Based on scale variations, an $\alpha$ shape uncertainty is determined, which is less than 2\% in any single \mT bin.
The JES and JER uncertainties are propagated to $\alpha$, and are not larger than 3\%.  
A summary of all systematic uncertainties related to the nonresonant background prediction is shown in 
the second and third columns of Table~\ref{tab:systsumBG}.

In addition to those listed in Table~\ref{tab:systsumBG}, we account for uncertainties due to the limited size of simulated data sets.
The Poisson uncertainties associated with the observed CR yields are propagated to the nonresonant background predictions in each \mT bin.
The uncertainties due to the limited size of simulated data sets are treated as uncorrelated across each analysis bin.

The predicted resonant background and signal yield uncertainties are evaluated from a larger list of potential sources.
These sources include the integrated luminosity, $\tau_{21}$ scale factors, pileup modeling, \PQb jet veto efficiency, effects related to
inefficiencies due to instrumental effects of the electromagnetic calorimeter trigger (prefiring), 
modeling of unclustered energy in the calculation of \ptmiss, jet mass scale (JMS) and
resolution (JMR), JES and JER, trigger efficiency modeling, PDF uncertainties, and scale uncertainties.  A summary of
the impacts these uncertainties have on resonant backgrounds and signal is shown in the fourth through seventh columns of
Table~\ref{tab:systsumBG} and in Table~\ref{tab:systsumSig}.  A description of each uncertainty source is provided below.

The measured integrated luminosity uncertainty is propagated to the prediction of the resonant background and signal yields.  This uncertainty amounts to
1.6\% on the entire data set~\cite{Sirunyan:2021qkt,CMS-PAS-LUM-17-004,CMS-PAS-LUM-18-002}
and is correlated across all signal regions.  The effects of the luminosity uncertainty on signal and nonresonant backgrounds are treated as fully correlated.

Uncertainties associated with the determination of $\tau_{21}$ efficiency scale factors that correct for systematic differences
in simulated and collision data sets are propagated to the predicted signal and resonant background yields. 
These uncertainties have two components: one that affects the normalization of predictions ($\tau_{21}$ SF) and 
another that accounts for \pt-dependent differences ($\tau_{21}$ \pt extrap.) in the tagging efficiency, which affects the \mT shape of our predicted yields.

Jet energy and jet mass scales are varied within their uncertainties.
Effects of JMS uncertainties are treated as anticorrelated between SR and CR regions and correlated across all categories.
The magnitudes of JMS uncertainties are assumed to be independent of \mT. They are treated as fully correlated across all \mT bins and all regions. 

Jet masses are smeared to broaden their distribution within measured JMR uncertainties. 
Jet mass smearing is done for both SR and CR but for simulated data sets only. The effect of jet smearing was at most 10\%. 
The effects of JMR uncertainties are correlated across \mT bins and all categories, and between signals and resonant backgrounds.

Similarly to the nonresonant background, the effect of JES and JER uncertainties are propagated to the predicted event yields.
These are found to have minimal impact on the \mT shape ($<$1\%) within a given category but can cause events to migrate from the ggF/DY
to the VBF categories.  

Other systematic uncertainties affect the normalization of resonant backgrounds and signals.  These include pileup uncertainties,
\PQb-tagging scale factor uncertainties, prefiring corrections, unclustered energy scale uncertainties,
and trigger uncertainties.  
These effects are assumed to be correlated across various categories, and between signal and resonant backgrounds. 

Finally, the statistical uncertainties due to the limited size of simulated data sets are propagated to all predicted signal 
and resonant background yields. In this analysis, all the uncertainties quoted are pre-fit values. 

\begin{table*}[!htbp]
\centering
\topcaption{Summary of systematic uncertainties (in \%) related to the SM background predictions in various regions.  
  Columns two and three tabulate the representative size of effects on $\alpha$ in the VBF and ggF/DY events categories,
  respectively.  Columns four through seven tabulate the typical size of effects on the prediction of resonant background 
  yields in the VBF SR, VBF CR, ggF/DY SR, and ggF/DY CR, respectively. All of these numbers are the pre-fit values. For 
  some systematic uncertainties, the variation in different \mT bins are shown as a range. Values of LP that are different from 
  those of HP are shown in parentheses.}
\begin{scotch}{lcccccc}
Source                                      & $\alpha_{\text{VBF}}$  & $\alpha_{\text{ggF/DY}}$    &          VBF SR  &         VBF CR        &          ggF/DY SR &       ggF/DY CR       \\
\hline
Luminosity                                  & \NA                    & \NA                         &           1.6    &         1.6           &     1.6            &       1.6            \\
$\tau_{21}$ SF                              & \NA                    & \NA                         &      7.3 (17.0)  &         7.3 (17.0)    &    7.3 (17.0)      &       7.3 (17.0)     \\
$\tau_{21}$ \pt extrap.                     & \NA                    & \NA                         &  2--18 (1--10)   &  2--18 (1--10)        &   2--18 (1--10)    &  2--18 (1--10)       \\
Pileup                                      & \NA                    & \NA                         &    3.9 (3.9)     &         2.0 (3.7)     &          1.0 (0.8) &       1.7 (0.9)      \\
\PQb jet veto                               & \NA                    & \NA                         &    2.4 (2.9)     &         3.5 (3.2)     &          1.8 (2.2) &       2.8 (2.6)      \\
Prefiring                                   & \NA                    & \NA                         &    0.7 (0.6)     &         0.8 (0.7)     &          0.3 (0.2) &       0.4 (0.3)      \\
Unclustered energy                             & \NA                    & \NA                         &      2.4 (1.6)   &         1.8 (1.4)     &      1.8 (1.3)     &       1.6 (1.5)      \\
JMS                                         & \NA                    & \NA                         &    0.5 (0.4)     &    1.8 (0.5)          &    0.3 (0.3)       &       1.6 (0.4)      \\
JMR                                         & \NA                    & \NA                         &    1.2 (1.6)     &         5.9 (0.7)     &      1.7 (1.01)    &       7.1 (0.96)     \\
Trigger                                     & \NA                    & \NA                         &     1.4          &      1.4              &     1.4            &   1.4                \\
JES                                         & 3.0                    & 1--2                        & 40--13           &  40--13               &    4.0             &       4.0            \\
JER                                         & 3.0                    & 1.5                         & 35--13           &  35--13               &    2.0             &       2.0            \\
PDF norm.                                   & \NA                    & \NA                         &    5.0           &         5.0           &    2.0             &       2.0            \\
PDF shape                                   & 3.0                    & 1.5                         &  0.5--4          & 0.5--4                &    0.5--4          &  0.5--4              \\
$\mu_\mathrm{R},\ \mu_\mathrm{F}$ norm.     & \NA                    & \NA                         &      15          &     15                &          11        &       12             \\
$\mu_\mathrm{R},\ \mu_\mathrm{F}$ shape     & 1--2                   & 1--2                        &  1--10           &  2--4                 &  3--8              &  3--4                \\
Nonclosure                                  & 7--40                  & 7--40                       &           \NA    &          \NA          &           \NA      &         \NA          \\
\end{scotch}
\label{tab:systsumBG}
\end{table*}

\begin{table*}[!htbp]
\centering
\topcaption{Summary of the typical size of systematic uncertainties (in \%) in the predicted signal yields in various regions.
        All of these numbers are the pre-fit values. A range is given for the shape systematic uncertainties. Values of LP that are different from those of HP are shown in parentheses.}
\begin{scotch}{lcccc}
Source                             &       VBF SR       &  VBF CR       &   ggF/DY SR   &   ggF/DY CR      \\
\hline
Luminosity                         &       1.6          &  1.6          &   1.6         &   1.6       \\
$\tau_{21}$ SF                     &       7.3 (17.0)   &  7.3 (17.0)   &   7.3 (17.0)  &   7.3 (17.0)     \\
$\tau_{21}$ \pt extrap.            &   2--18 (1--10)    & 2--18 (1--10) & 2--18 (1--10) & 2--18 (1--10)    \\
Pileup                             &       0.2 (0.5)    &  0.7 (1.0)    &   1.1 (0.9)   &   1.1 (0.6)      \\
\PQb jet veto                         &       1.3 (1.4)    &  1.4 (1.6)    &   1.3 (1.4)   &   1.3 (1.5)      \\
Prefiring                          &       1.0 (0.9)    &  1.0 (1.0)    &   0.6 (0.5)   &   0.6 (0.5)      \\
Unclustered energy                    &       0.1          &  0.1          &   0.1         &   0.1       \\
JMS                                &       0.9 (0.3)    &  2.6 (1.9)    &   0.7 (0.3)   &   2.4 (1.5)      \\
JMR                                &       3.2 (2.8)    &  6.9 (4.0)    &   3.3 (2.4)   &   7.5 (3.4)      \\
Trigger                            &       1.4 (1.4)    &  1.4 (1.5)    &   1.4 (1.6)   &   1.4 (1.5)      \\
JES                                & 5--11              & 4--14         & 1--7    & 1--5       \\
JER                                & 2--3               & 6--7          & 1--7    & 1--5       \\
\end{scotch}
\label{tab:systsumSig}
\end{table*}

\section{Results and interpretations}
\label{sec:results}

The final predicted event yields are computed using a simultaneous maximum likelihood fit of event yields in all \mT bins,
and all SRs and CRs. 
Each bin is modeled as a marked Poisson model~\cite{VISCHIA2020100046}
with mean value corresponding to the sum of expected yields for resonant and nonresonant backgrounds, and signal.
An unconstrained nuisance parameter is implemented to allow for the fit to independently adjust the nonresonant background in each \mT bin of each category.
For each \mT bin of each event category, this constraint is fully correlated between the SR and CR. Systematic uncertainties are implemented using log-normal priors.
The likelihood is parameterized in terms of the signal strength $\mu$, which is the ratio of the measured signal cross section and the theoretical cross section.

\begin{figure*}[!htbp]
  \centering
    \includegraphics[width=0.45\textwidth]{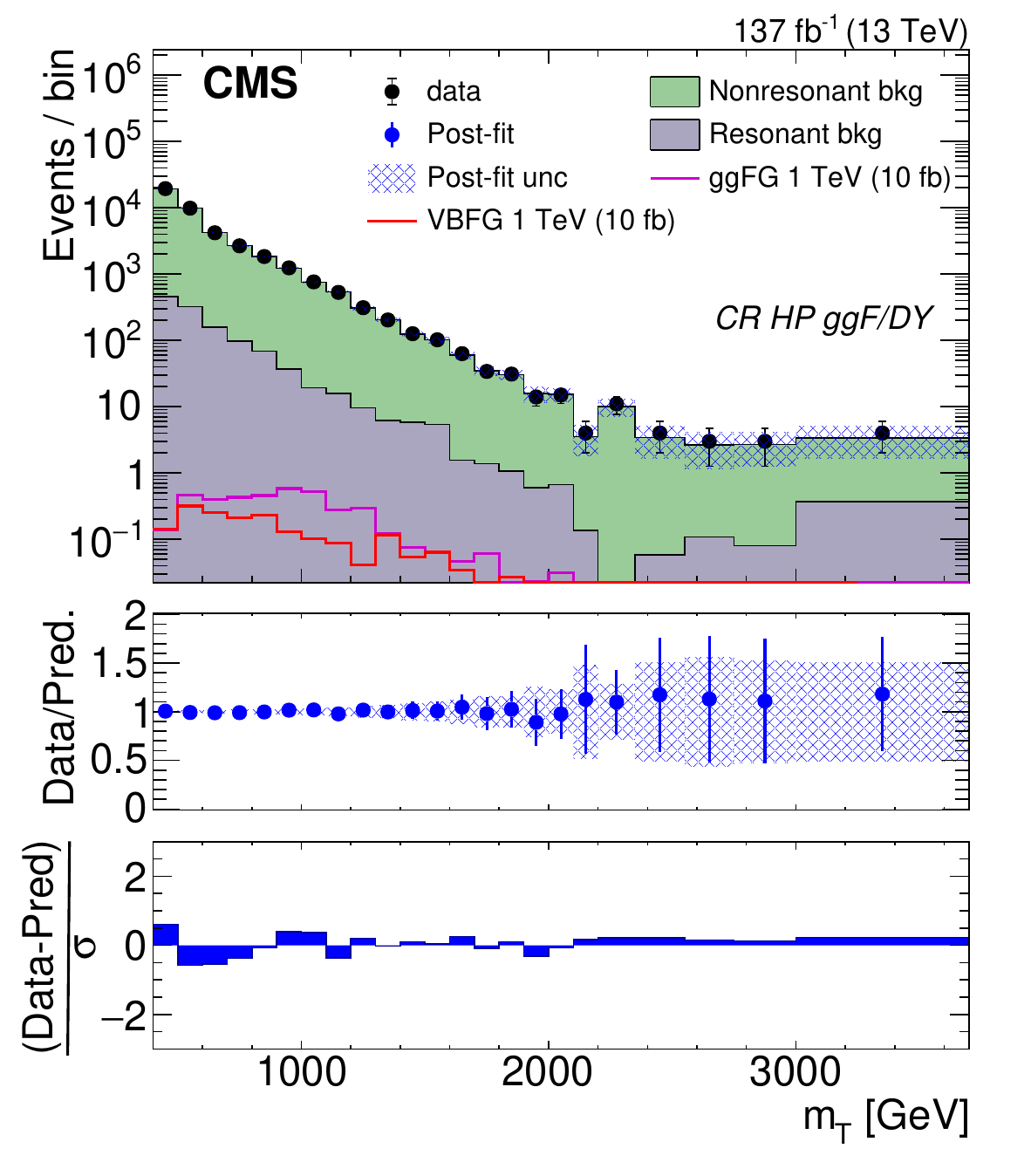}
    \includegraphics[width=0.45\textwidth]{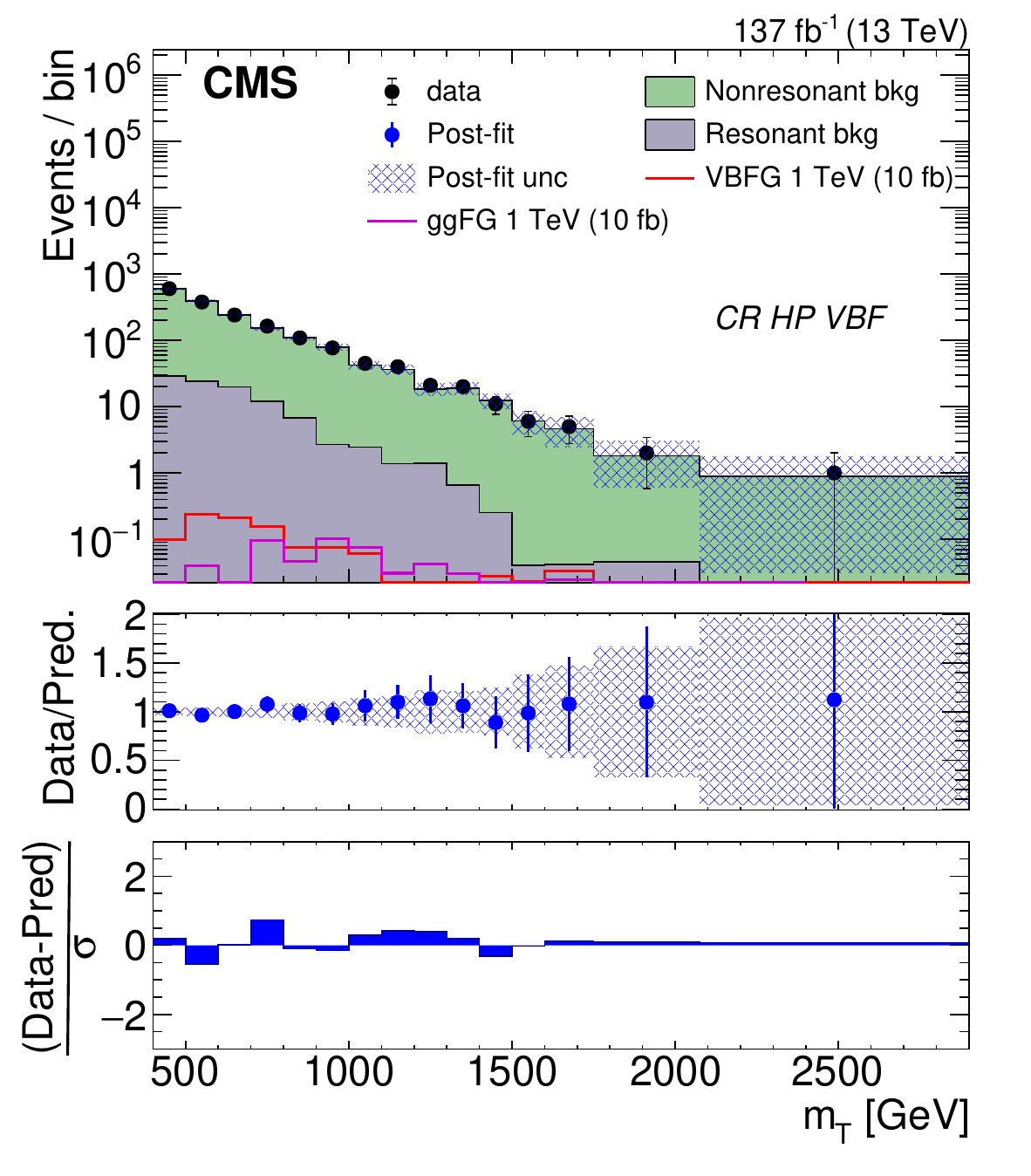} \\
    \includegraphics[width=0.45\textwidth]{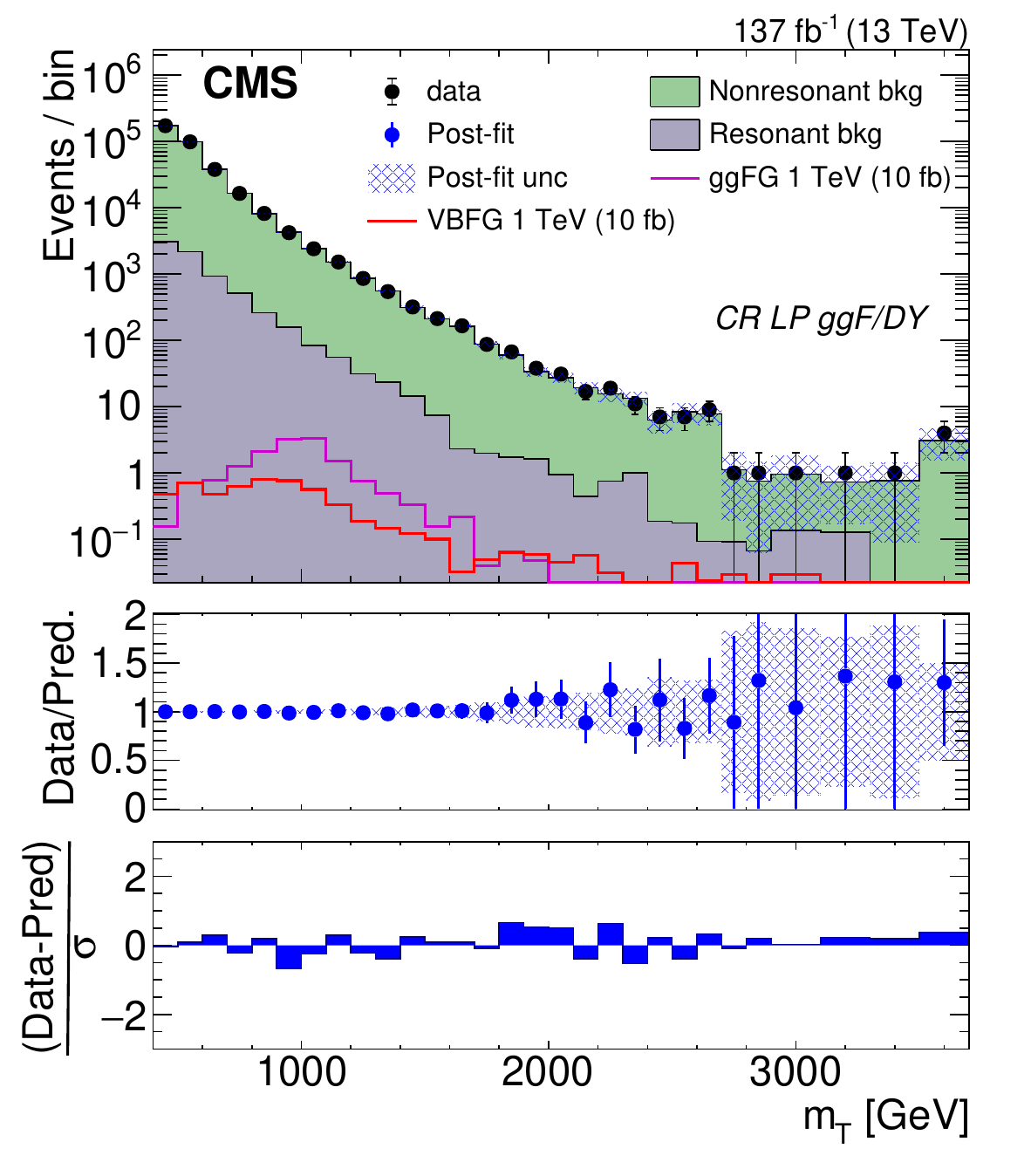}
    \includegraphics[width=0.45\textwidth]{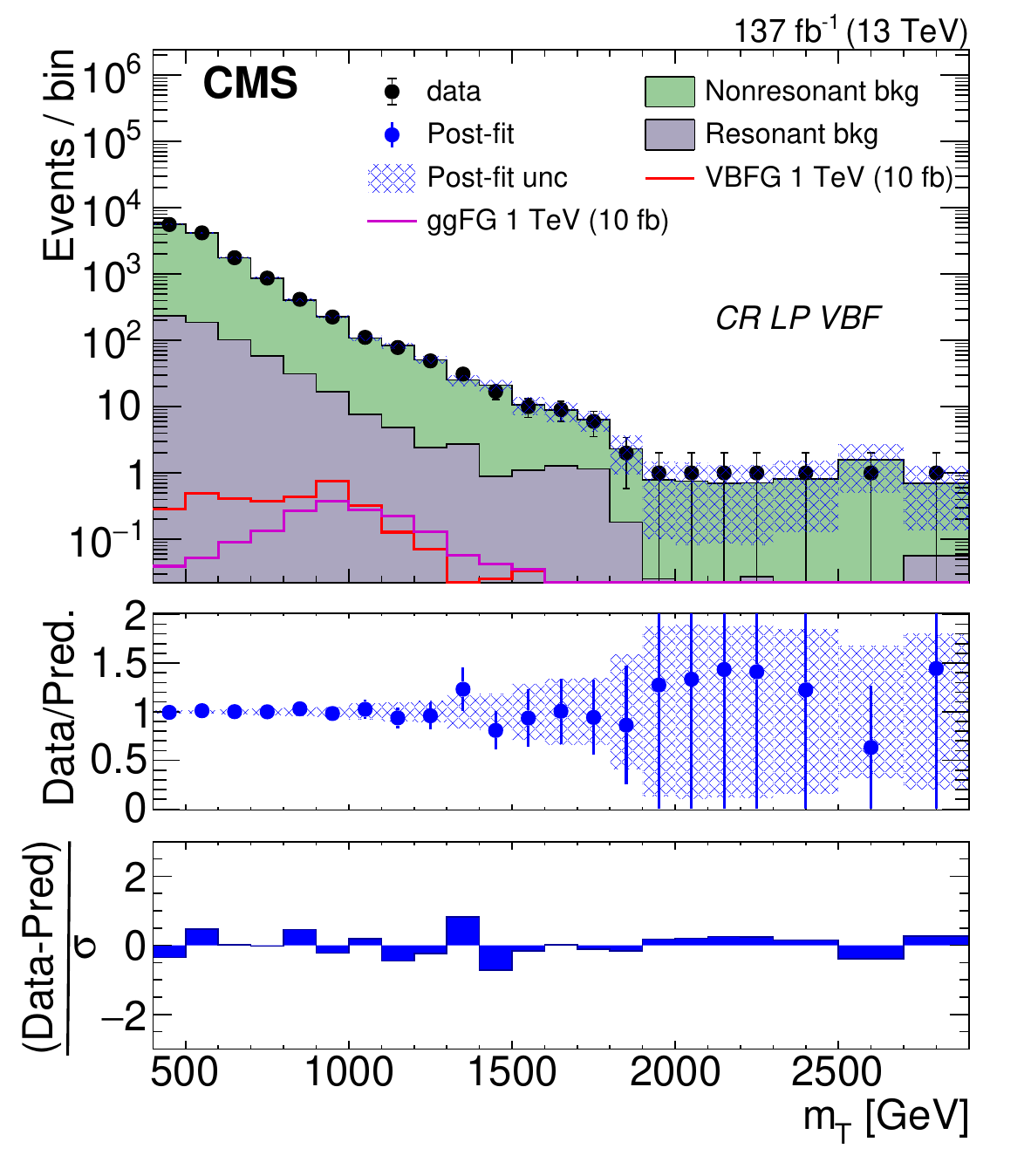} \\

  \caption{Distributions of \mT for high-purity ggF/DY (upper left) and VBF (upper right), and 
    low-purity ggF/DY (lower left) and VBF (lower right) CR events after performing background-only fits. 
    The last bin in the upper left, upper right, lower left, and lower right plot corresponds to the yields integrated above 3, 2.3, 3.5, and 2.7\TeV, respectively.
    The top panel of each plot shows the post-fit prediction, represented by filled histograms, compared to observed yields, represented by black points.  
    Both the ggF and VBF-produced 1\TeV graviton signals are shown in each plot, represented by the open purple and red histograms, respectively. 
    The signal is normalized to 10\unit{fb}. The blue hashed area represents  
    the total uncertainty from the post-fit predicted event yield as a function of \mT. The middle panel of each plot shows the ratio of data and post-fit predictions in blue. 
    The bottom panel of each plot shows the difference between the observed event yields and the post-fit predictions normalized by the quadratic sum of the
    statistical uncertainty of the observed yield and the total uncertainty from the post-fit prediction in each \mT bin.}
  \label{fig:Bkg_MT_Run2_HP}
\end{figure*}

\begin{figure*}[!htbp]
  \centering
    \includegraphics[width=0.45\textwidth]{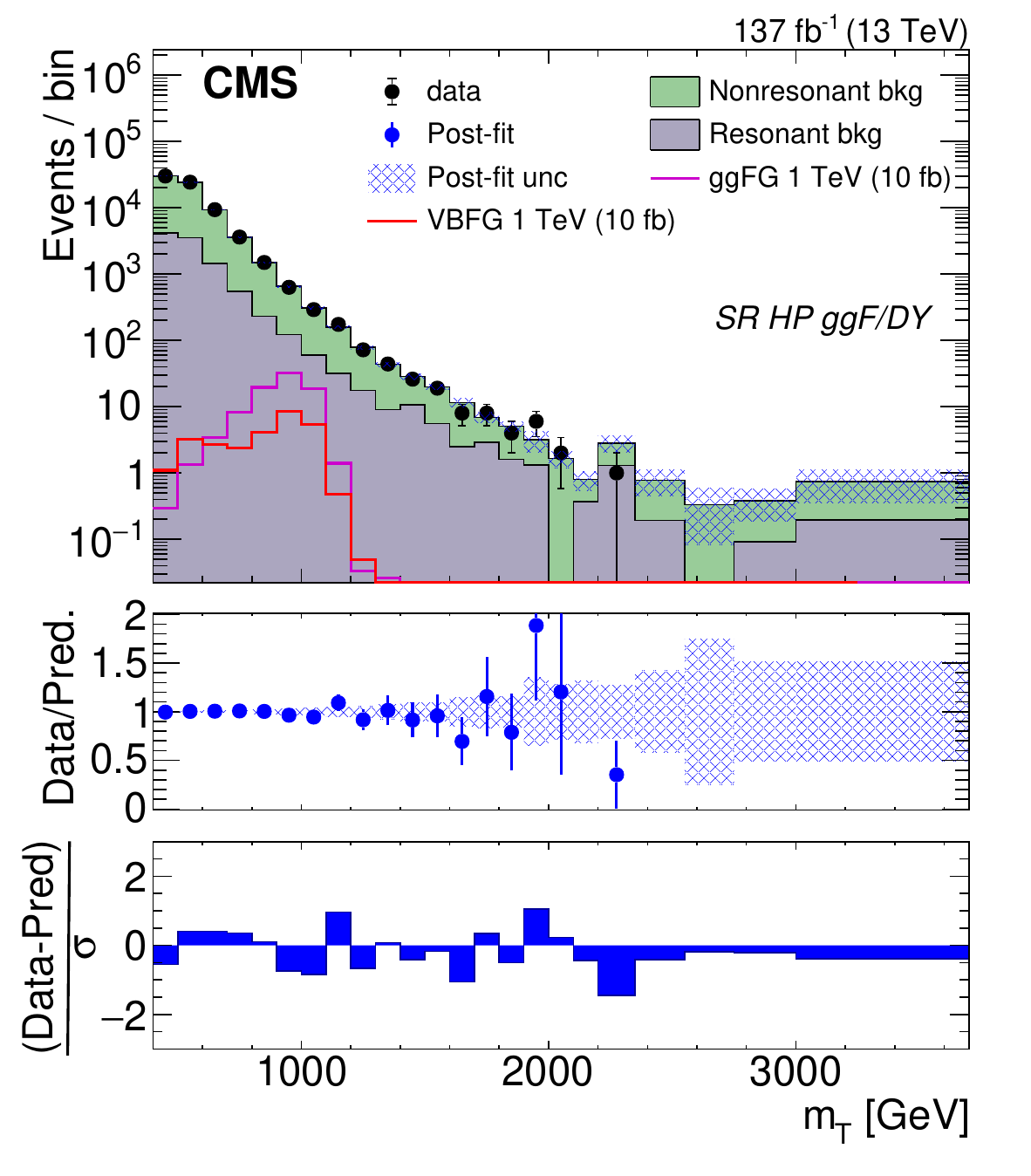}
    \includegraphics[width=0.45\textwidth]{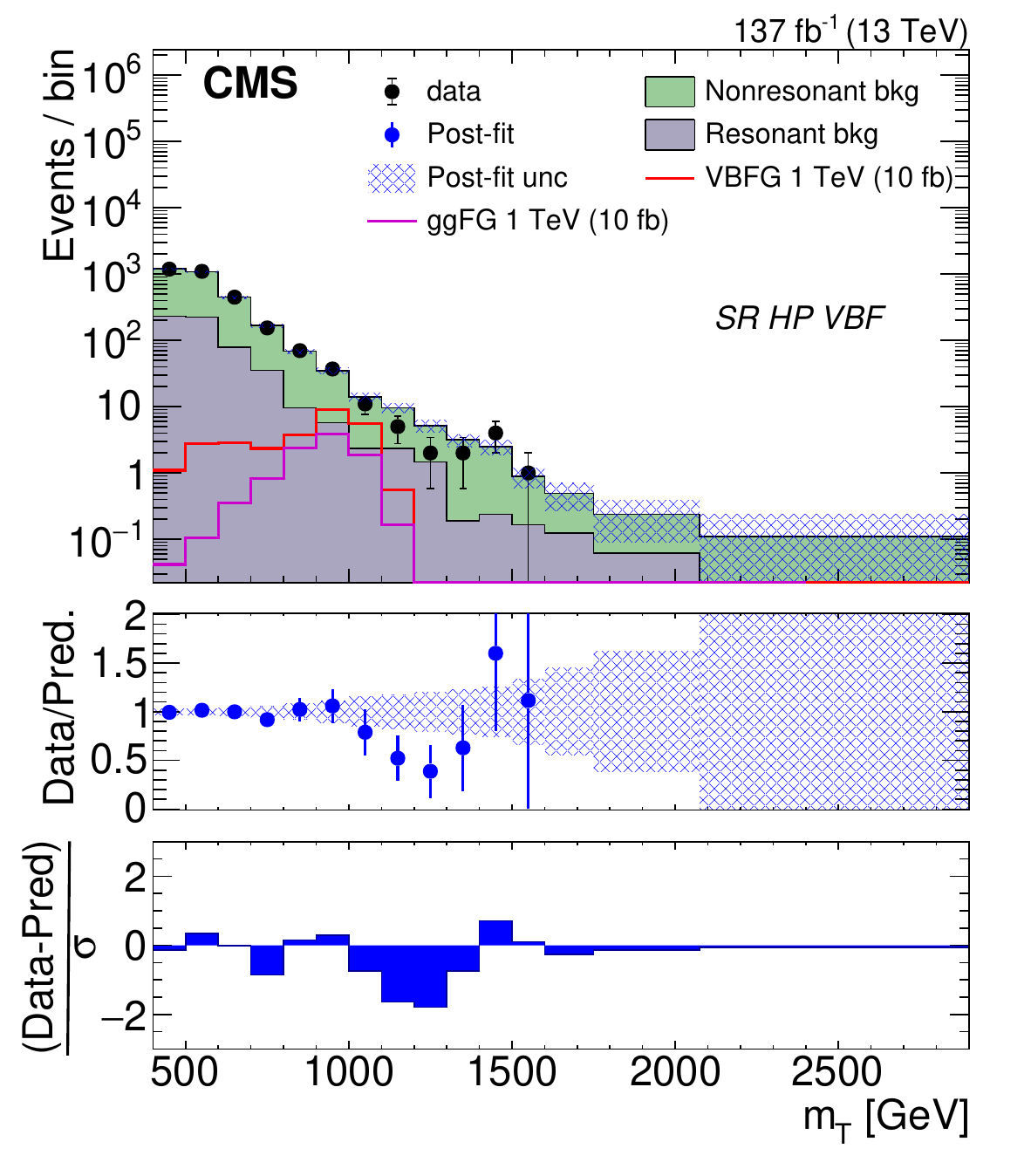}\\
    \includegraphics[width=0.45\textwidth]{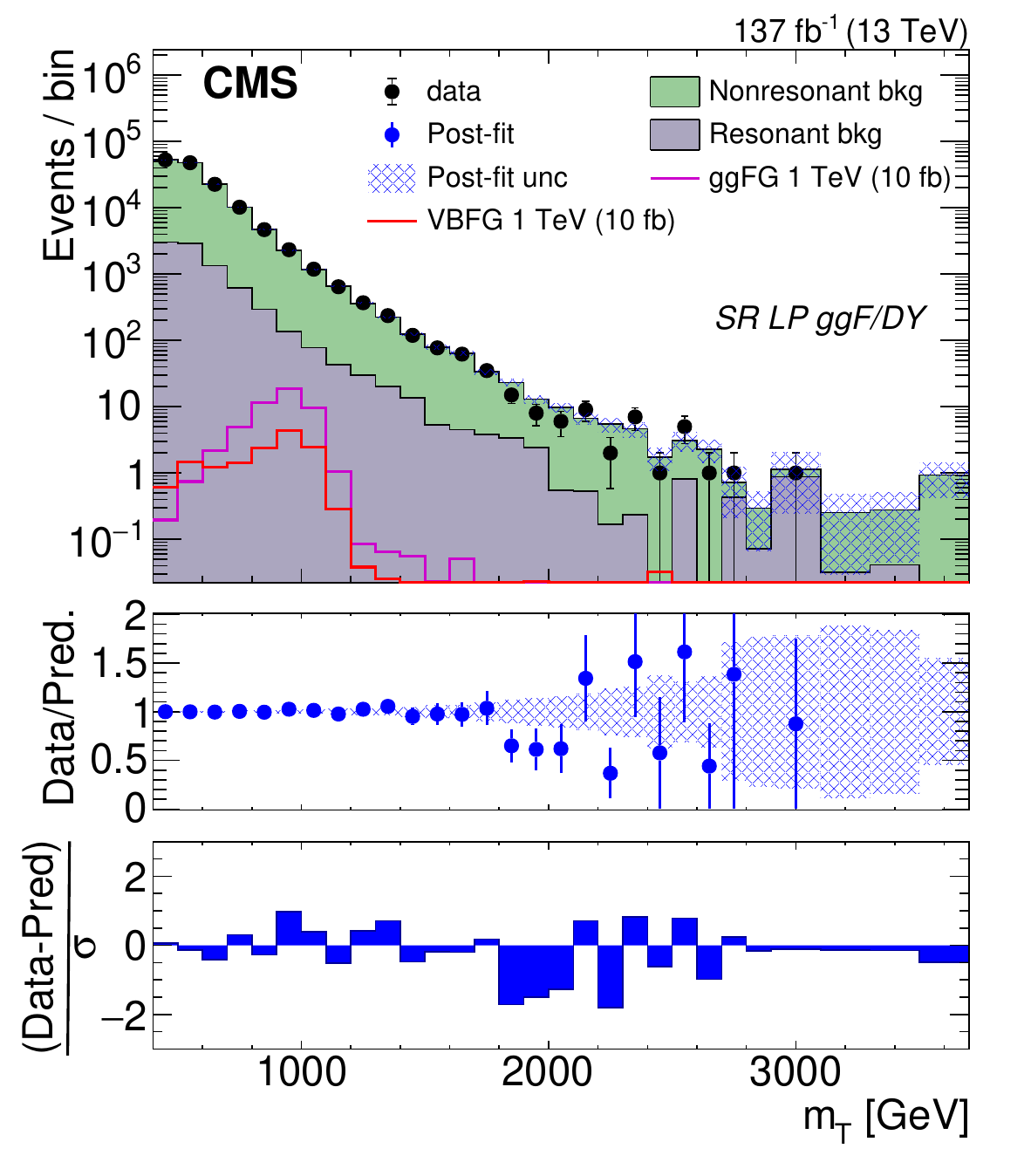}
    \includegraphics[width=0.45\textwidth]{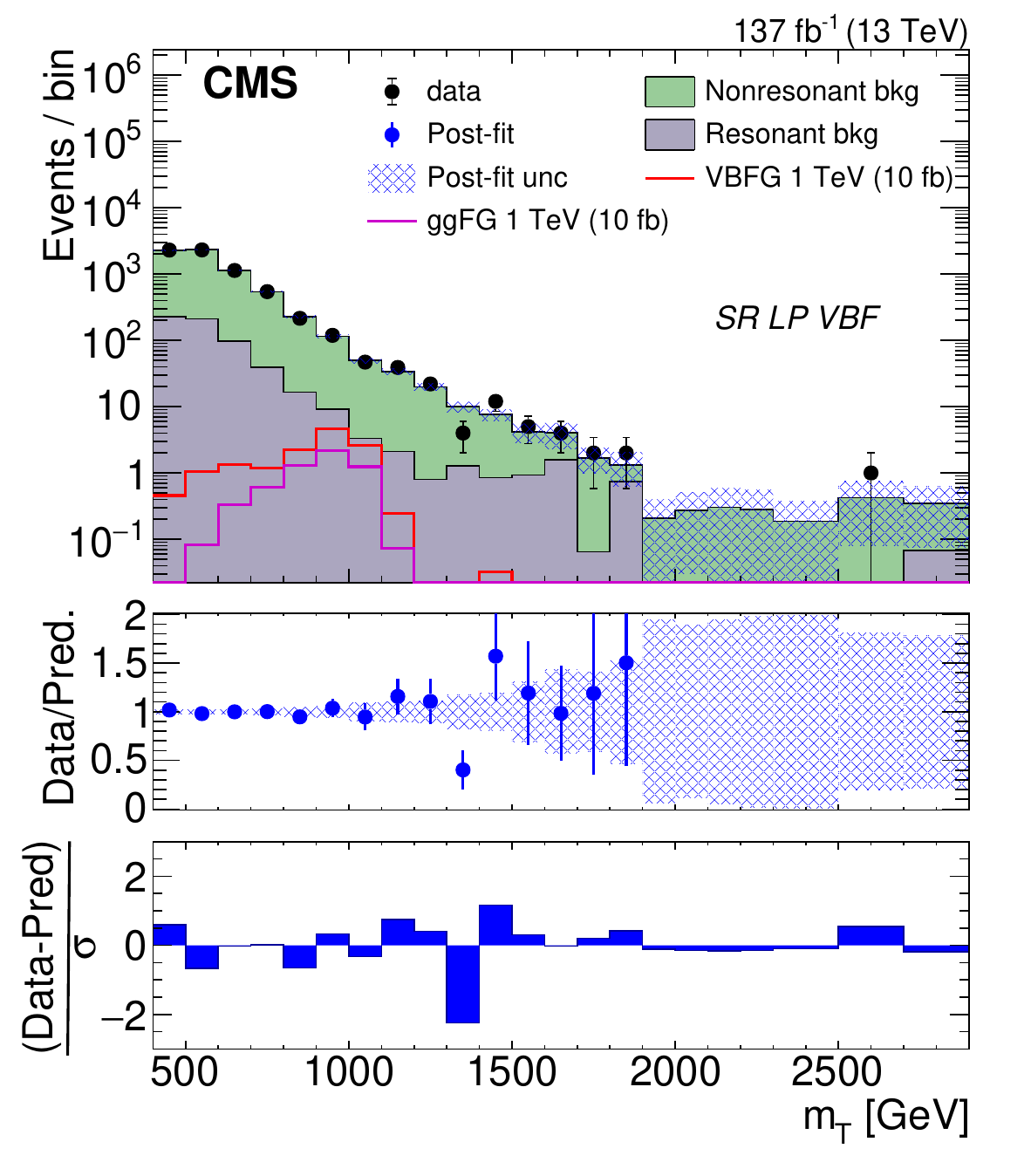}\\

    \caption{Distribution of the predicted and observed event yields versus \mT for high-purity ggF/DY (upper left) and VBF (upper right), and 
      low-purity ggF/DY (lower left) and VBF (lower right) SR events. 
      The last bin in each plot corresponds to the yields integrated above the penultimate bin. 
      The top panel of each plot shows the prediction based on a background-only fit to data, represented by filled histograms, 
      compared to observed yields, represented by black points.  
      Both the ggF and VBF-produced 1\TeV graviton signals are shown in each plot, represented by the open purple and red histograms, respectively. 
      The signal is normalized to 10\unit{fb}. The middle panel of each plot shows the ratio of data and post-fit predictions in blue. 
      The blue hashed area represents the total uncertainty from the post-fit predicted event yield as a function of \mT.
      The bottom panel of each plot shows the difference between the observed event yields and the post-fit predictions normalized by the quadratic sum of the
      statistical uncertainty of the observed yield and the total uncertainty from the post-fit prediction in each \mT bin.}
  \label{fig:pred_MT_Run2_HP}
\end{figure*}

The predicted event yields for all backgrounds and a graviton ($m_{\PXXG}=1\TeV$) are shown in 
Fig.~\ref{fig:Bkg_MT_Run2_HP} (\ref{fig:pred_MT_Run2_HP}) for the CR (SR). 
The data are compared to post-fit predictions, where fit refers to constraints on predictions and their uncertainties based on a
maximum likelihood fit to data in which the signal strength is fixed to $\mu=0$. 
The post-fit predictions and observations are consistent within the uncertainties, suggesting no evidence of new sources of diboson production.

We derive both expected and observed 95\% confidence level (\CL) upper limits on the 
{$\PX\to\PV(\PQq\PAQq)\PZ(\PGn\PAGn)$} production cross section.
A test statistic is used in conjunction with the \CLs criterion~\cite{bib-cls} to set upper limits.  The test statistic is defined as 
$q_{\mu}=-2\ln(\mathcal{L}_{\mu}/\mathcal{L}_{\text{max}})$, where $\
\mathcal{L}_{\text{max}}$ refers to the maximum value of the likelihood when all parameters are varied and $\mathcal{L}_{\mu}$ refers to the likelihood 
obtained by varying $\mu$ while profiling all the other parameters conditioned on its value.  Upper limits
are computed using the asymptotic approximation~\cite{Cowan:2010js}.
Expected limits are computed by evaluating the test statistic using the post-fit predicted numbers of background events and their uncertainties.

\begin{figure}[!htb]
  \centering
    \includegraphics[width=0.45\textwidth]{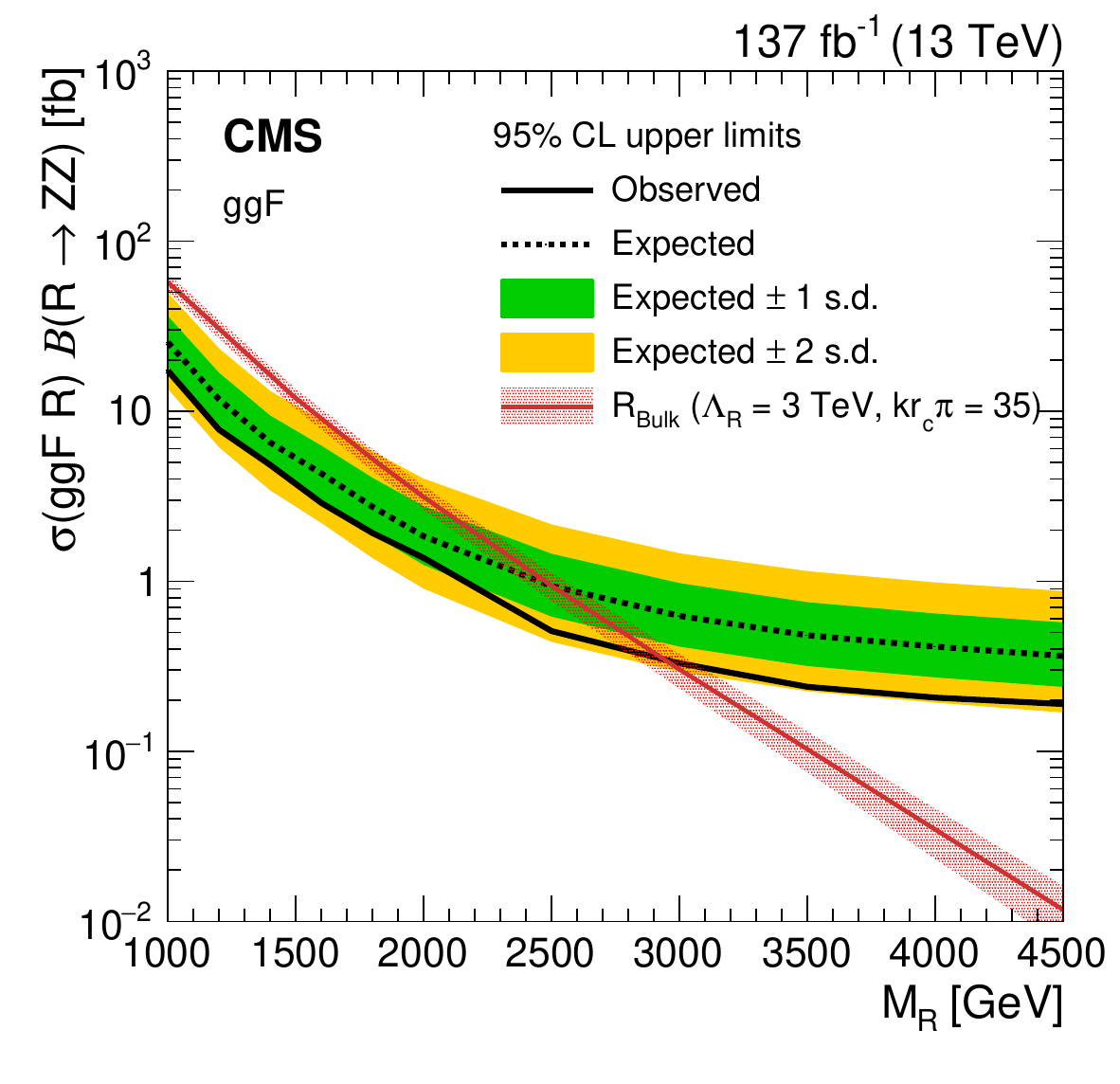}
    \includegraphics[width=0.45\textwidth]{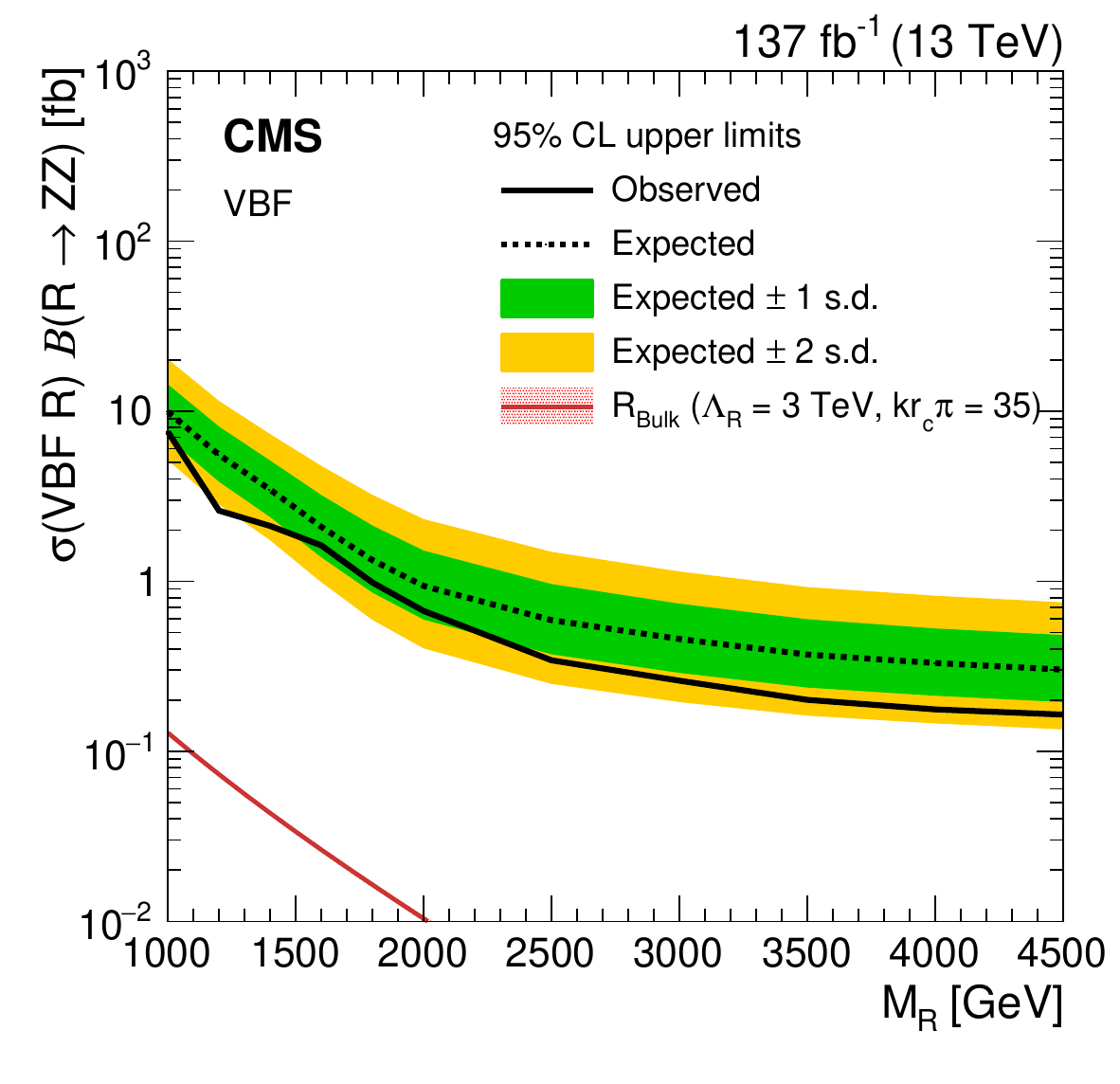}
    \caption{Expected and observed 95\% \CL upper limits on the radion (\PR) production cross section times
      the $\PR\to \PZ\PZ$ branching fraction versus the radion mass are shown as dashed and solid black lines, respectively.
      Green and yellow bands, respectively, represent the 68\% and 95\% confidence intervals of the expected limit.
      The red curves show the theoretical radion production cross sections times their branching fractions to \PZ{}\PZ.
      The hashed red areas represent the theoretical cross section uncertainty due to limited knowledge of PDFs and scale choices.
      Limits and theory cross sections for ggF-produced radions are shown in the \cmsLeft figure, 
      while the \cmsRight figure shows the same for VBF-produced radions.}
  \label{fig:radion_limits}
\end{figure}

\begin{figure}[!htb]
  \centering
    \includegraphics[width=0.45\textwidth]{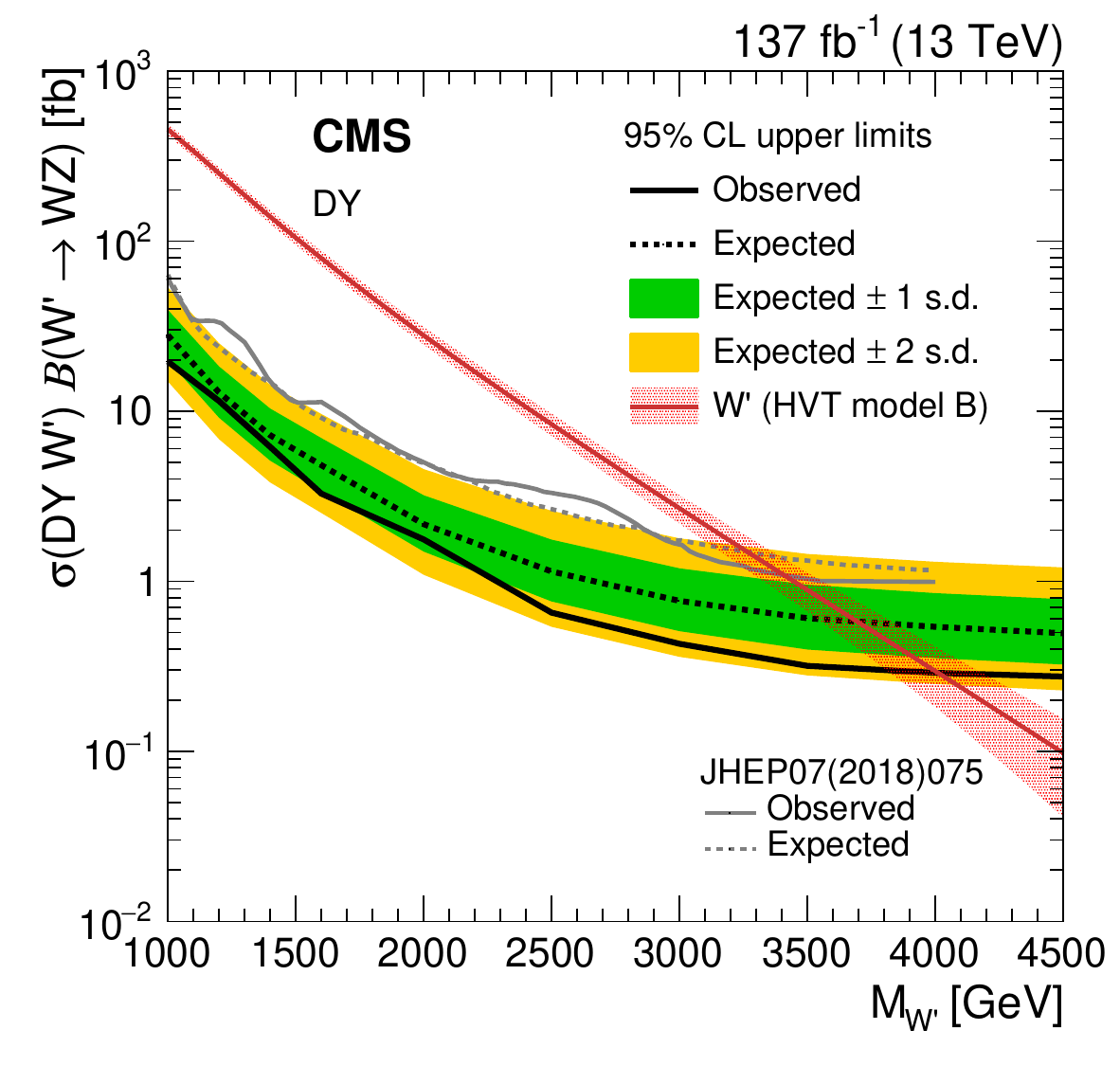}
    \includegraphics[width=0.45\textwidth]{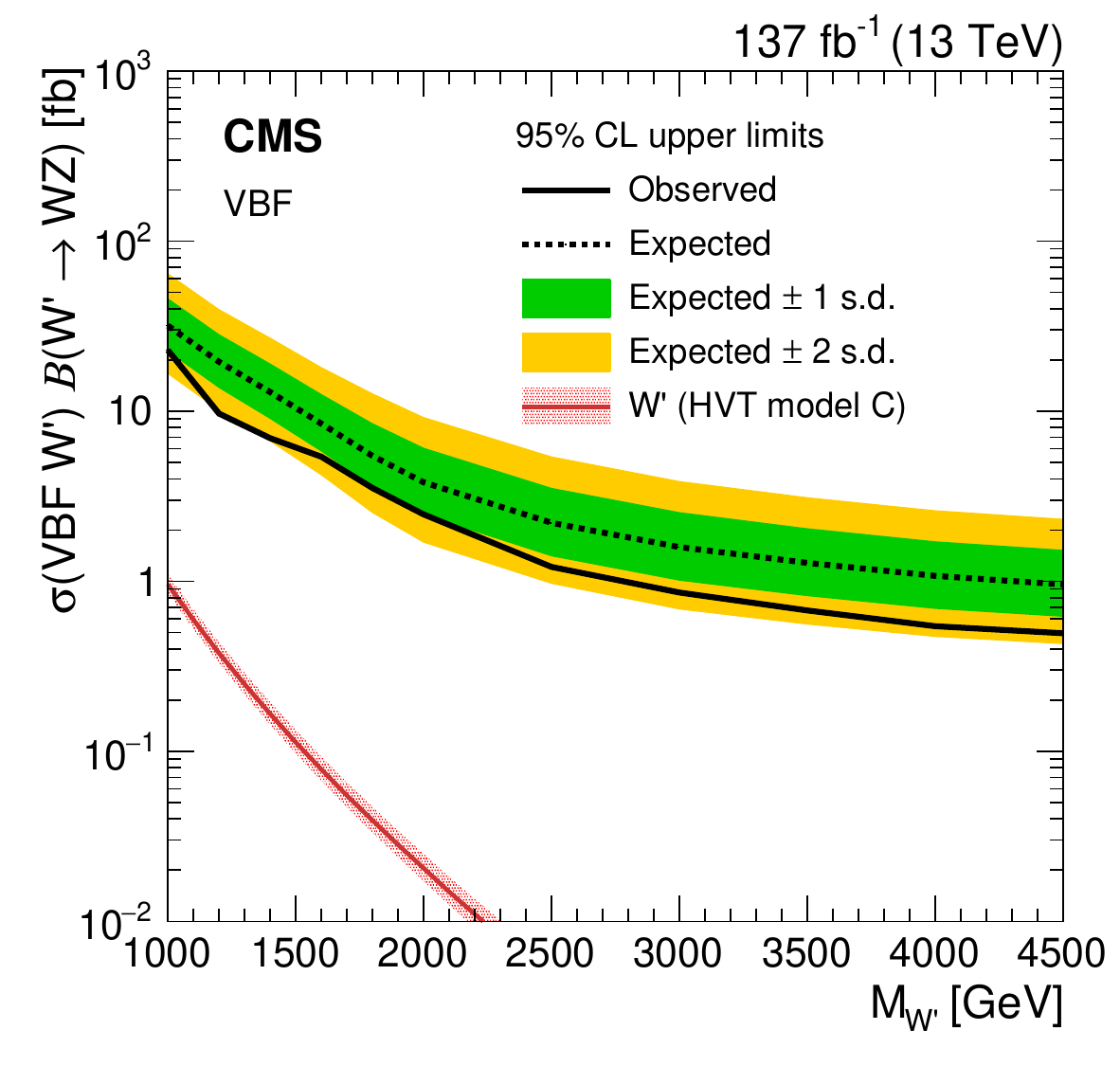}
  \caption{Expected and observed 95\% \CL upper limits on the \PWpr production cross section
    times the $\PWpr\to \PW\PZ$ branching fraction versus the \PWpr mass are shown as dashed and solid black lines, respectively.
    Green and yellow bands, respectively, represent the 68\% and 95\% confidence intervals of the expected limit.
    The red curves show the theoretical \PWpr boson production cross sections times their branching fractions to \PW{}\PZ.
    The hashed red areas represent the theoretical cross section uncertainty due to limited knowledge of PDFs and scale choices.
    Limits and theory cross sections for DY-produced \PWpr bosons are shown in the \cmsLeft
    figure, while the \cmsRight figure shows the same for VBF-produced \PWpr bosons. The grey curves in the \cmsLeft plot show the previous CMS results with 36\fbinv of data.}
  \label{fig:Wprime_limits}
\end{figure}

\begin{figure}[!htb]
  \centering
    \includegraphics[width=0.45\textwidth]{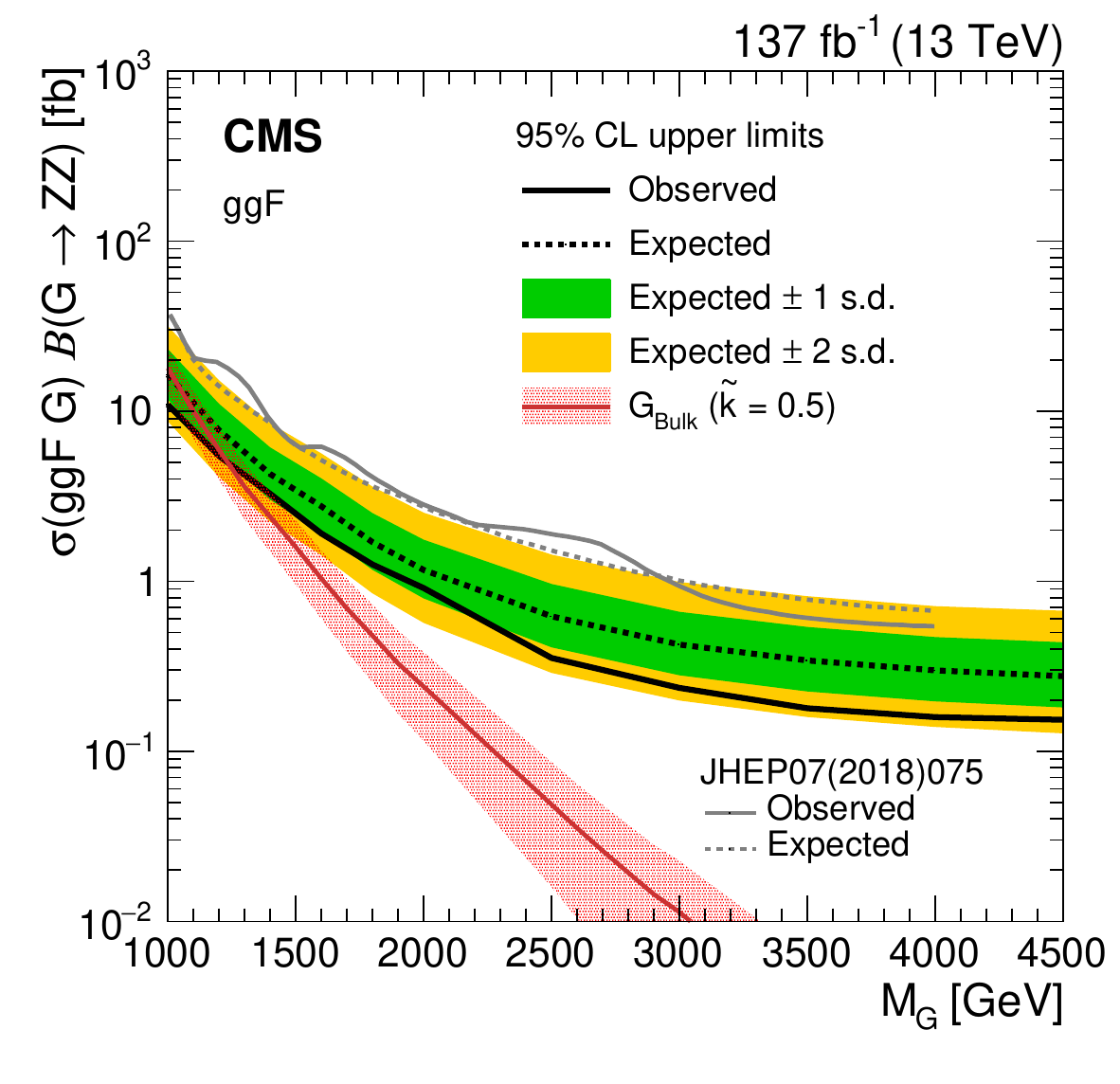}
    \includegraphics[width=0.45\textwidth]{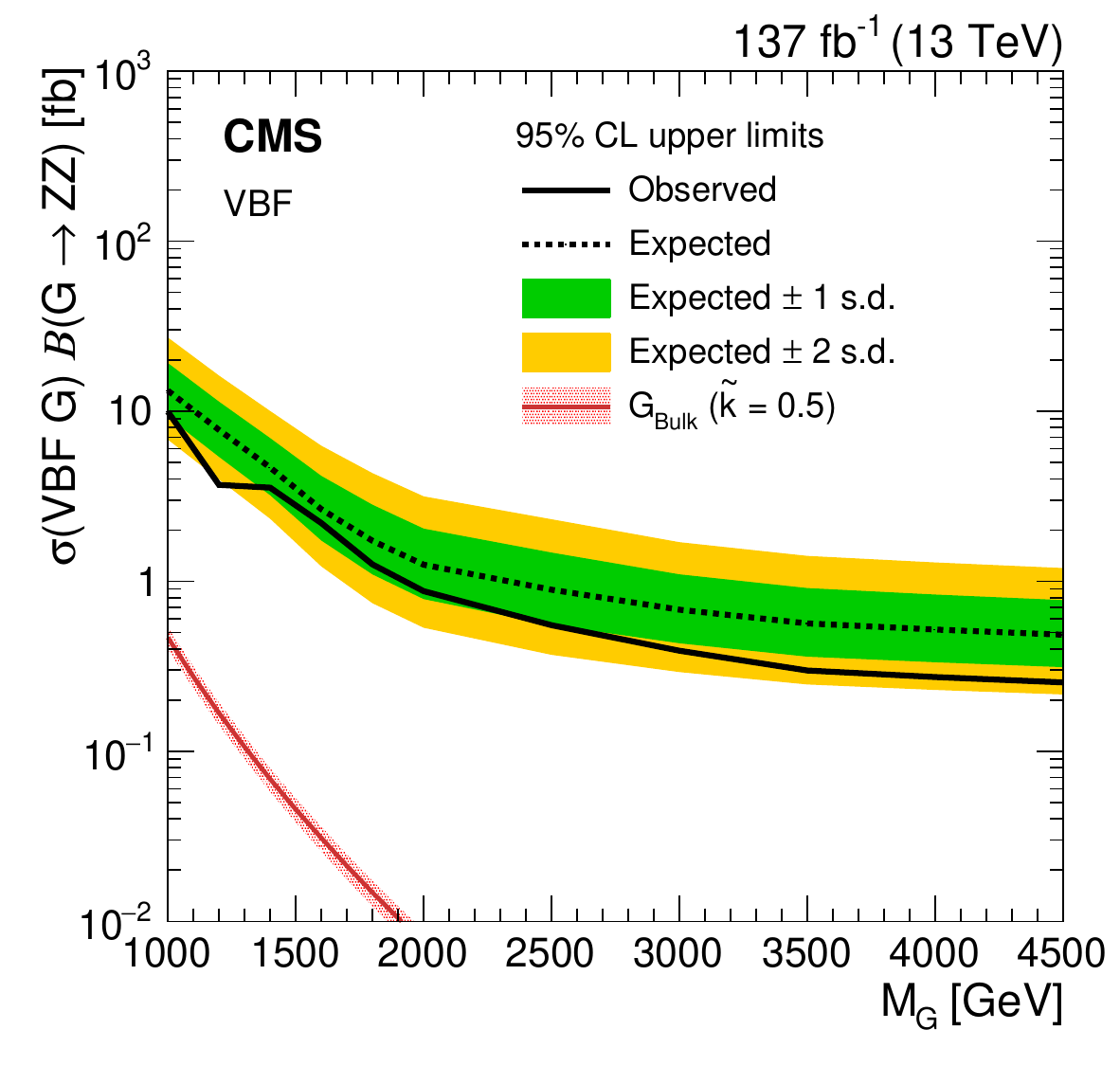}
  \caption{Expected and observed 95\% \CL upper limits on the graviton (\PXXG) production cross section
    times the $\PXXG\to \PZ\PZ$ branching fraction versus the graviton mass are shown as dashed and solid black lines, respectively.
    Green and yellow bands, respectively, represent 68\% and 95\% confidence intervals of the expected limit.
    The red curves show the theoretical graviton production cross sections times their branching fractions to \PZ{}\PZ.
    The hashed red areas represent the theoretical cross section uncertainty due to limited knowledge of PDFs and scale choices.
    Limits and theory cross sections for ggF-produced gravitons are shown in the \cmsLeft figure, while the \cmsRight figure shows 
    the same for VBF-produced gravitons. The grey curves in the \cmsLeft plot show the previous CMS results with 36\fbinv of data.}
  \label{fig:graviton_limits}
\end{figure}

Upper limits on the radion production cross sections times their branching fraction to \PZ{}\PZ versus the radion mass 
are shown in Fig.~\ref{fig:radion_limits}.
Limits are computed assuming radions are produced entirely either through the ggF process or the VBF process.  
The expected (observed) radion mass exclusion limits are 2.5 (3.0)\TeV for ggF-produced states.
These are the first mass exclusion limits set by CMS on ggF-produced radions in this final state.  
Figure~\ref{fig:Wprime_limits} shows the expected and observed upper limits on the \PWpr boson production cross sections times their branching fraction 
to \PW{}\PZ, assuming exclusive production through the DY (model B) or VBF process (model C).
The expected and observed mass exclusion limits for DY-produced \PWpr resonances are found to be 3.7 and 4.0\TeV, respectively. 
This is an improvement of 0.6 (0.4)\TeV in the observed (expected) mass exclusion limit from the previous CMS result.
Finally, the upper limits on the graviton production cross sections times their branching fraction to \PZ{}\PZ are shown in Fig.~\ref{fig:graviton_limits}.
Limits are set assuming gravitons are produced entirely either through the ggF process or the VBF process.  
The expected (observed) graviton mass exclusion limit is found to be 1.1 (1.2)\TeV, assuming gravitons are produced exclusively through the ggF mechanism.
These are the first mass exclusion limits set by CMS on ggF-produced gravitons in this final state.  
For the VBF-produced models considered here, no masses are excluded.
The observed upper limits on $\sigma{\mathcal{B}(\PX\to\PV\PZ)}$  vary between 0.2 and 9\unit{fb} for radions, 0.5 and 
20\unit{fb} for \PWpr resonances, and 0.3 and 10\unit{fb} for gravitons.  

The methods used here complement the recent ATLAS search~\cite{ATLAS:2020fry} in the same channel by using different jet substructure variables and different VBF tagging requirements. 
The 95\% \CL upper limits on these resonance production cross sections times $\PX\to \PZ+\PW/\PZ$ branching 
fraction from the recent ATLAS results are comparable to those set in this paper.

\section{Summary}
\label{sec:summary}

A search has been presented for new bosonic states decaying either to a pair of \PZ bosons or to a \PW boson and a \PZ boson.
The analyzed final states require large missing transverse momentum and one high-momentum, large-radius jet.
Large-radius jets are required to have a mass consistent with either a \PW or \PZ boson.
Events are categorized based on the presence of large-radius jets passing high-purity and low-purity substructure requirements.
Events are also categorized based on the presence or absence of high-momentum jets in the forward region of the detector.
Forward jets distinguish weak vector boson fusion (VBF) from other production mechanisms.
Contributions from the dominant SM backgrounds are estimated from data control regions using an extrapolation method.
No deviation between SM expectation and data is found, and 95\% confidence level upper limits are set on the
production cross section times branching fraction for several signal models.
A lower observed (expected) limit of 3.0 (2.5)\TeV is set on the mass of gluon-gluon fusion produced radions.
The observed (expected) mass exclusion limit for Drell--Yan produced \PWpr bosons is found to be 4.0 (3.7)\TeV. 
The observed (expected) mass exclusion limit for gluon-gluon fusion produced gravitons is found to be 1.2 (1.1)\TeV.
At 95\% confidence level, upper observed (expected) limits on the VBF production cross section 
times $\PX\to \PZ+\PW/\PZ$ branching fraction range between 0.2 and 20 (0.3 and 30)\unit{fb}. 
The 95\% \CL upper limits on these resonance production cross sections times $\PX\to \PZ+\PW/\PZ$ branching 
fraction from the recent ATLAS results are comparable to those set in this paper.

\begin{acknowledgments}
  We congratulate our colleagues in the CERN accelerator departments for the excellent performance of the LHC and thank the technical and administrative staffs at CERN and at other CMS institutes for their contributions to the success of the CMS effort. In addition, we gratefully acknowledge the computing centers and personnel of the Worldwide LHC Computing Grid and other centers for delivering so effectively the computing infrastructure essential to our analyses. Finally, we acknowledge the enduring support for the construction and operation of the LHC, the CMS detector, and the supporting computing infrastructure provided by the following funding agencies: BMBWF and FWF (Austria); FNRS and FWO (Belgium); CNPq, CAPES, FAPERJ, FAPERGS, and FAPESP (Brazil); MES and BNSF (Bulgaria); CERN; CAS, MoST, and NSFC (China); MINCIENCIAS (Colombia); MSES and CSF (Croatia); RIF (Cyprus); SENESCYT (Ecuador); MoER, ERC PUT and ERDF (Estonia); Academy of Finland, MEC, and HIP (Finland); CEA and CNRS/IN2P3 (France); BMBF, DFG, and HGF (Germany); GSRI (Greece); NKFIA (Hungary); DAE and DST (India); IPM (Iran); SFI (Ireland); INFN (Italy); MSIP and NRF (Republic of Korea); MES (Latvia); LAS (Lithuania); MOE and UM (Malaysia); BUAP, CINVESTAV, CONACYT, LNS, SEP, and UASLP-FAI (Mexico); MOS (Montenegro); MBIE (New Zealand); PAEC (Pakistan); MSHE and NSC (Poland); FCT (Portugal); JINR (Dubna); MON, RosAtom, RAS, RFBR, and NRC KI (Russia); MESTD (Serbia); SEIDI, CPAN, PCTI, and FEDER (Spain); MOSTR (Sri Lanka); Swiss Funding Agencies (Switzerland); MST (Taipei); ThEPCenter, IPST, STAR, and NSTDA (Thailand); TUBITAK and TAEK (Turkey); NASU (Ukraine); STFC (United Kingdom); DOE and NSF (USA).
    
  \hyphenation{Rachada-pisek} Individuals have received support from the Marie-Curie program and the European Research Council and Horizon 2020 Grant, contract Nos.\ 675440, 724704, 752730, 758316, 765710, 824093, 884104, and COST Action CA16108 (European Union); the Leventis Foundation; the Alfred P.\ Sloan Foundation; the Alexander von Humboldt Foundation; the Belgian Federal Science Policy Office; the Fonds pour la Formation \`a la Recherche dans l'Industrie et dans l'Agriculture (FRIA-Belgium); the Agentschap voor Innovatie door Wetenschap en Technologie (IWT-Belgium); the F.R.S.-FNRS and FWO (Belgium) under the ``Excellence of Science -- EOS" -- be.h project n.\ 30820817; the Beijing Municipal Science \& Technology Commission, No. Z191100007219010; the Ministry of Education, Youth and Sports (MEYS) of the Czech Republic; the Deutsche Forschungsgemeinschaft (DFG), under Germany's Excellence Strategy -- EXC 2121 ``Quantum Universe" -- 390833306, and under project number 400140256 - GRK2497; the Lend\"ulet (``Momentum") Program and the J\'anos Bolyai Research Scholarship of the Hungarian Academy of Sciences, the New National Excellence Program \'UNKP, the NKFIA research grants 123842, 123959, 124845, 124850, 125105, 128713, 128786, and 129058 (Hungary); the Council of Science and Industrial Research, India; the Latvian Council of Science; the Ministry of Science and Higher Education and the National Science Center, contracts Opus 2014/15/B/ST2/03998 and 2015/19/B/ST2/02861 (Poland); the Funda\c{c}\~ao para a Ci\^encia e a Tecnologia, grant CEECIND/01334/2018 (Portugal); the National Priorities Research Program by Qatar National Research Fund; the Ministry of Science and Higher Education, project no. 14.W03.31.0026 (Russia); the Programa Estatal de Fomento de la Investigaci{\'o}n Cient{\'i}fica y T{\'e}cnica de Excelencia Mar\'{\i}a de Maeztu, grant MDM-2015-0509 and the Programa Severo Ochoa del Principado de Asturias; the Stavros Niarchos Foundation (Greece); the Rachadapisek Sompot Fund for Postdoctoral Fellowship, Chulalongkorn University and the Chulalongkorn Academic into Its 2nd Century Project Advancement Project (Thailand); the Kavli Foundation; the Nvidia Corporation; the SuperMicro Corporation; the Welch Foundation, contract C-1845; and the Weston Havens Foundation (USA).
\end{acknowledgments}

\bibliography{auto_generated}   

\cleardoublepage \appendix\section{The CMS Collaboration \label{app:collab}}\begin{sloppypar}\hyphenpenalty=5000\widowpenalty=500\clubpenalty=5000\input{B2G-20-008-authorlist.tex}\end{sloppypar}
\end{document}

%% file: B2G-20-008-authorlist.tex
\vskip\cmsinstskip
\textbf{Yerevan Physics Institute, Yerevan, Armenia}\\*[0pt]
A.~Tumasyan
\vskip\cmsinstskip
\textbf{Institut f\"{u}r Hochenergiephysik, Vienna, Austria}\\*[0pt]
W.~Adam, J.W.~Andrejkovic, T.~Bergauer, S.~Chatterjee, M.~Dragicevic, A.~Escalante~Del~Valle, R.~Fr\"{u}hwirth\cmsAuthorMark{1}, M.~Jeitler\cmsAuthorMark{1}, N.~Krammer, L.~Lechner, D.~Liko, I.~Mikulec, P.~Paulitsch, F.M.~Pitters, J.~Schieck\cmsAuthorMark{1}, R.~Sch\"{o}fbeck, M.~Spanring, S.~Templ, W.~Waltenberger, C.-E.~Wulz\cmsAuthorMark{1}
% ...
\vskip\cmsinstskip
% ...
1:  Also at TU Wien, Wien, Austria\\
% ...


%% file: B2G-20-008_temp.bbl
\providecommand{\href}[2]{#2}\begingroup\raggedright\begin{thebibliography}{10}%
\makeatletter
\providecommand{\hrefCMSnoop }[0]{\@secondoftwo}%
\makeatother
\providecommand{\doi}{\texttt{doi:}\begingroup \urlstyle{tt}\Url}

\bibitem{Randall_1999a}
\hrefCMSnoop {}{L.~Randall and R.~Sundrum, ``Large mass hierarchy from a small
  extra dimension'',} \textit{ Phys. Rev. Lett.} \textbf{ 83} (1999) 3370,
  \href{http://dx.doi.org/10.1103/physrevlett.83.3370}{\doi{10.1103/physrevlett.83.3370}},
  \href{http://www.arXiv.org/abs/hep-ph/9905221}{\texttt{arXiv:hep-ph/9905221}}.

\bibitem{Randall_1999b}
\hrefCMSnoop {}{L.~Randall and R.~Sundrum, ``An alternative to
  compactification'',} \textit{ Phys. Rev. Lett.} \textbf{ 83} (1999) 4690,
  \href{http://dx.doi.org/10.1103/physrevlett.83.4690}{\doi{10.1103/physrevlett.83.4690}},
  \href{http://www.arXiv.org/abs/hep-th/9906064}{\texttt{arXiv:hep-th/9906064}}.

\bibitem{Agashe:2008jb}
K.~Agashe\hrefCMSnoop {}{ {et~al.}, ``{LHC} signals for warped electroweak
  charged gauge bosons'',} \textit{ Phys. Rev. D} \textbf{ 80} (2009) 075007,
  \href{http://dx.doi.org/10.1103/PhysRevD.80.075007}{\doi{10.1103/PhysRevD.80.075007}},
  \href{http://www.arXiv.org/abs/0810.1497}{\texttt{arXiv:0810.1497}}.

\bibitem{Agashe:2009bb}
K.~Agashe\hrefCMSnoop {}{ {et~al.}, ``{LHC} signals for coset electroweak gauge
  bosons in warped/composite {pseudo-Goldstone} boson {Higgs} models'',}
  \textit{ Phys. Rev. D} \textbf{ 81} (2010) 096002,
  \href{http://dx.doi.org/10.1103/PhysRevD.81.096002}{\doi{10.1103/PhysRevD.81.096002}},
  \href{http://www.arXiv.org/abs/0911.0059}{\texttt{arXiv:0911.0059}}.

\bibitem{Agashe:2007ki}
K.~Agashe\hrefCMSnoop {}{ {et~al.}, ``{CERN} {LHC} signals for warped
  electroweak neutral gauge bosons'',} \textit{ Phys. Rev. D} \textbf{ 76}
  (2007) 115015,
  \href{http://dx.doi.org/10.1103/PhysRevD.76.115015}{\doi{10.1103/PhysRevD.76.115015}},
  \href{http://www.arXiv.org/abs/0709.0007}{\texttt{arXiv:0709.0007}}.

\bibitem{Pappadopulo:2014qza}
\hrefCMSnoop {}{D.~Pappadopulo, A.~Thamm, R.~Torre, and A.~Wulzer, ``Heavy
  vector triplets: Bridging theory and data'',} \textit{ JHEP} \textbf{ 09}
  (2014) 060,
  \href{http://dx.doi.org/10.1007/JHEP09(2014)060}{\doi{10.1007/JHEP09(2014)060}},
  \href{http://www.arXiv.org/abs/1402.4431}{\texttt{arXiv:1402.4431}}.

\bibitem{ArkaniHamed:2002qx}
N.~Arkani-Hamed\hrefCMSnoop {}{ {et~al.}, ``The minimal moose for a little
  {Higgs}'',} \textit{ JHEP} \textbf{ 08} (2002) 021,
  \href{http://dx.doi.org/10.1088/1126-6708/2002/08/021}{\doi{10.1088/1126-6708/2002/08/021}},
  \href{http://www.arXiv.org/abs/hep-ph/0206020}{\texttt{arXiv:hep-ph/0206020}}.

\bibitem{ArkaniHamed:2002qy}
\hrefCMSnoop {}{N.~Arkani-Hamed, A.~G. Cohen, E.~Katz, and A.~E. Nelson, ``The
  littlest {Higgs}'',} \textit{ JHEP} \textbf{ 07} (2002) 034,
  \href{http://dx.doi.org/10.1088/1126-6708/2002/07/034}{\doi{10.1088/1126-6708/2002/07/034}},
  \href{http://www.arXiv.org/abs/hep-ph/0206021}{\texttt{arXiv:hep-ph/0206021}}.

\bibitem{Burdman:2002ns}
\hrefCMSnoop {}{G.~Burdman, M.~Perelstein, and A.~Pierce, ``{Large Hadron
  Collider} tests of the little {Higgs} model'',} \textit{ Phys. Rev. Lett.}
  \textbf{ 90} (2003) 241802,
  \href{http://dx.doi.org/10.1103/PhysRevLett.90.241802}{\doi{10.1103/PhysRevLett.90.241802}},
  \href{http://www.arXiv.org/abs/hep-ph/0212228}{\texttt{arXiv:hep-ph/0212228}}.
  [Erratum: \DOI{10.1103/PhysRevLett.92.049903}].

\bibitem{Agashe:2007zd}
\hrefCMSnoop {}{K.~Agashe, H.~Davoudiasl, G.~Perez, and A.~Soni, ``Warped
  gravitons at the {CERN} {LHC} and beyond'',} \textit{ Phys. Rev. D} \textbf{
  76} (2007) 036006,
  \href{http://dx.doi.org/10.1103/PhysRevD.76.036006}{\doi{10.1103/PhysRevD.76.036006}},
  \href{http://www.arXiv.org/abs/hep-ph/0701186}{\texttt{arXiv:hep-ph/0701186}}.

\bibitem{Fitzpatrick:2007qr}
\hrefCMSnoop {}{A.~L. Fitzpatrick, J.~Kaplan, L.~Randall, and L.-T. Wang,
  ``Searching for the {Kaluza-Klein} graviton in bulk {RS} models'',} \textit{
  JHEP} \textbf{ 09} (2007) 013,
  \href{http://dx.doi.org/10.1088/1126-6708/2007/09/013}{\doi{10.1088/1126-6708/2007/09/013}},
  \href{http://www.arXiv.org/abs/hep-ph/0701150}{\texttt{arXiv:hep-ph/0701150}}.

\bibitem{Aaboud:2017itg}
\hrefCMSnoop {}{{ATLAS Collaboration}, ``{Searches for heavy $ZZ$ and $ZW$
  resonances in the $\ell\ell qq$ and $\nu\nu qq$ final states in $pp$
  collisions at $\sqrt{s}=13$ TeV with the {ATLAS} detector}'',} \textit{ JHEP}
  \textbf{ 03} (2018) 009,
  \href{http://dx.doi.org/10.1007/JHEP03(2018)009}{\doi{10.1007/JHEP03(2018)009}},
  \href{http://www.arXiv.org/abs/1708.09638}{\texttt{arXiv:1708.09638}}.

\bibitem{Sirunyan:2018ivv}
\hrefCMSnoop {}{{CMS Collaboration}, ``{Search for a heavy resonance decaying
  into a Z boson and a vector boson in the $ \nu
  \overline{\nu}\mathrm{q}\overline{\mathrm{q}} $ final state}'',} \textit{
  JHEP} \textbf{ 07} (2018) 075,
  \href{http://dx.doi.org/10.1007/JHEP07(2018)075}{\doi{10.1007/JHEP07(2018)075}},
  \href{http://www.arXiv.org/abs/1803.03838}{\texttt{arXiv:1803.03838}}.

\bibitem{ATLAS:2020fry}
\hrefCMSnoop {}{{ATLAS Collaboration}, ``{Search for heavy diboson resonances
  in semileptonic final states in pp collisions at $\sqrt{s}=13$ TeV with the
  ATLAS detector}'',} \textit{ Eur. Phys. J. C} \textbf{ 80} (2020) 1165,
  \href{http://dx.doi.org/10.1140/epjc/s10052-020-08554-y}{\doi{10.1140/epjc/s10052-020-08554-y}},
  \href{http://www.arXiv.org/abs/2004.14636}{\texttt{arXiv:2004.14636}}.

\bibitem{hepdata}
\hrefCMSnoop {}{``{HEPD}ata record for this analysis'',} 2021.
\newblock
  \href{http://dx.doi.org/10.17182/hepdata.103856}{\doi{10.17182/hepdata.103856}}.

\bibitem{Sirunyan:2020zal}
\hrefCMSnoop {}{{CMS Collaboration}, ``{Performance of the {CMS} Level-1
  trigger in proton-proton collisions at $\sqrt{s} = 13$\,{TeV}}'',} \textit{
  JINST} \textbf{ 15} (2020) P10017,
  \href{http://dx.doi.org/10.1088/1748-0221/15/10/P10017}{\doi{10.1088/1748-0221/15/10/P10017}},
  \href{http://www.arXiv.org/abs/2006.10165}{\texttt{arXiv:2006.10165}}.

\bibitem{Khachatryan:2016bia}
\hrefCMSnoop {}{{CMS Collaboration}, ``{The {CMS} trigger system}'',} \textit{
  JINST} \textbf{ 12} (2017) P01020,
  \href{http://dx.doi.org/10.1088/1748-0221/12/01/P01020}{\doi{10.1088/1748-0221/12/01/P01020}},
\href{http://www.arXiv.org/abs/1609.02366}{\texttt{arXiv:1609.02366}}.

\bibitem{Chatrchyan:2008zzk}
\hrefCMSnoop {}{{CMS Collaboration}, ``The {CMS} experiment at the {CERN}
  {LHC}'',} \textit{ JINST} \textbf{ 3} (2008) S08004,
  \href{http://dx.doi.org/10.1088/1748-0221/3/08/S08004}{\doi{10.1088/1748-0221/3/08/S08004}}.

\bibitem{Alwall:2014hca}
J.~Alwall\hrefCMSnoop {}{ {et~al.}, ``{The automated computation of tree-level
  and next-to-leading order differential cross sections, and their matching to
  parton shower simulations}'',} \textit{ JHEP} \textbf{ 07} (2014) 079,
  \href{http://dx.doi.org/10.1007/JHEP07(2014)079}{\doi{10.1007/JHEP07(2014)079}},
\href{http://www.arXiv.org/abs/1405.0301}{\texttt{arXiv:1405.0301}}.

\bibitem{Frederix:2012ps}
\hrefCMSnoop {}{R.~Frederix and S.~Frixione, ``{Merging meets matching in
  {MC@NLO}}'',} \textit{ JHEP} \textbf{ 12} (2012) 061,
  \href{http://dx.doi.org/10.1007/JHEP12(2012)061}{\doi{10.1007/JHEP12(2012)061}},
\href{http://www.arXiv.org/abs/1209.6215}{\texttt{arXiv:1209.6215}}.

\bibitem{Alwall:2007fs}
J.~Alwall\hrefCMSnoop {}{ {et~al.}, ``Comparative study of various algorithms
  for the merging of parton showers and matrix elements in hadronic
  collisions'',} \textit{ Eur. Phys. J. C} \textbf{ 53} (2008) 473,
  \href{http://dx.doi.org/10.1140/epjc/s10052-007-0490-5}{\doi{10.1140/epjc/s10052-007-0490-5}},
\href{http://www.arXiv.org/abs/0706.2569}{\texttt{arXiv:0706.2569}}.

\bibitem{Artoisenet:2012st}
\hrefCMSnoop {}{P.~Artoisenet, R.~Frederix, O.~Mattelaer, and R.~Rietkerk,
  ``Automatic spin-entangled decays of heavy resonances in {Monte Carlo}
  simulations'',} \textit{ JHEP} \textbf{ 03} (2013) 015,
  \href{http://dx.doi.org/10.1007/JHEP03(2013)015}{\doi{10.1007/JHEP03(2013)015}},
  \href{http://www.arXiv.org/abs/1212.3460}{\texttt{arXiv:1212.3460}}.

\bibitem{Nason:2004rx}
\hrefCMSnoop {}{P.~Nason, ``{A new method for combining {NLO} {QCD} with shower
  {Monte Carlo} algorithms}'',} \textit{ JHEP} \textbf{ 11} (2004) 040,
  \href{http://dx.doi.org/10.1088/1126-6708/2004/11/040}{\doi{10.1088/1126-6708/2004/11/040}},
\href{http://www.arXiv.org/abs/hep-ph/0409146}{\texttt{arXiv:hep-ph/0409146}}.

\bibitem{Frixione:2007vw}
\hrefCMSnoop {}{S.~Frixione, P.~Nason, and C.~Oleari, ``{Matching {NLO} {QCD}
  computations with parton shower simulations: the {POWHEG} method}'',}
  \textit{ JHEP} \textbf{ 11} (2007) 070,
  \href{http://dx.doi.org/10.1088/1126-6708/2007/11/070}{\doi{10.1088/1126-6708/2007/11/070}},
\href{http://www.arXiv.org/abs/0709.2092}{\texttt{arXiv:0709.2092}}.

\bibitem{Alioli:2010xd}
\hrefCMSnoop {}{S.~Alioli, P.~Nason, C.~Oleari, and E.~Re, ``{A general
  framework for implementing {NLO} calculations in shower {Monte Carlo}
  programs: the {POWHEG} {BOX}}'',} \textit{ JHEP} \textbf{ 06} (2010) 043,
  \href{http://dx.doi.org/10.1007/JHEP06(2010)043}{\doi{10.1007/JHEP06(2010)043}},
\href{http://www.arXiv.org/abs/1002.2581}{\texttt{arXiv:1002.2581}}.

\bibitem{Alioli:2009je}
\hrefCMSnoop {}{S.~Alioli, P.~Nason, C.~Oleari, and E.~Re, ``{{NLO} single-top
  production matched with shower in {POWHEG}: $s$- and $t$-channel
  contributions}'',} \textit{ JHEP} \textbf{ 09} (2009) 111,
  \href{http://dx.doi.org/10.1088/1126-6708/2009/09/111}{\doi{10.1088/1126-6708/2009/09/111}},
  \href{http://www.arXiv.org/abs/0907.4076}{\texttt{arXiv:0907.4076}}.
[Erratum: \DOI{10.1007/JHEP02(2010)011}].

\bibitem{Re:2010bp}
\hrefCMSnoop {}{E.~Re, ``{Single-top {Wt-channel} production matched with
  parton showers using the {POWHEG} method}'',} \textit{ Eur. Phys. J. C}
  \textbf{ 71} (2011) 1547,
  \href{http://dx.doi.org/10.1140/epjc/s10052-011-1547-z}{\doi{10.1140/epjc/s10052-011-1547-z}},
\href{http://www.arXiv.org/abs/1009.2450}{\texttt{arXiv:1009.2450}}.

\bibitem{Sjostrand:2014zea}
T.~Sj{\"o}strand\hrefCMSnoop {}{ {et~al.}, ``An introduction to {PYTHIA}
  8.2'',} \textit{ Comput. Phys. Commun.} \textbf{ 191} (2015) 159,
  \href{http://dx.doi.org/10.1016/j.cpc.2015.01.024}{\doi{10.1016/j.cpc.2015.01.024}},
\href{http://www.arXiv.org/abs/1410.3012}{\texttt{arXiv:1410.3012}}.

\bibitem{Khachatryan:2015pea}
\hrefCMSnoop {}{{CMS Collaboration}, ``{Event generator tunes obtained from
  underlying event and multiparton scattering measurements}'',} \textit{ Eur.
  Phys. J. C} \textbf{ 76} (2016) 155,
  \href{http://dx.doi.org/10.1140/epjc/s10052-016-3988-x}{\doi{10.1140/epjc/s10052-016-3988-x}},
\href{http://www.arXiv.org/abs/1512.00815}{\texttt{arXiv:1512.00815}}.

\bibitem{Sirunyan:2019dfx}
\hrefCMSnoop {}{{CMS Collaboration}, ``{Extraction and validation of a new set
  of {CMS} {PYTHIA8} tunes from underlying-event measurements}'',} \textit{
  Eur. Phys. J. C} \textbf{ 80} (2020) 4,
  \href{http://dx.doi.org/10.1140/epjc/s10052-019-7499-4}{\doi{10.1140/epjc/s10052-019-7499-4}},
  \href{http://www.arXiv.org/abs/1903.12179}{\texttt{arXiv:1903.12179}}.

\bibitem{Ball:2014uwa}
\hrefCMSnoop {}{{NNPDF} Collaboration, ``{Parton distributions for the {LHC}
  Run II}'',} \textit{ JHEP} \textbf{ 04} (2015) 040,
  \href{http://dx.doi.org/10.1007/JHEP04(2015)040}{\doi{10.1007/JHEP04(2015)040}},
  \href{http://www.arXiv.org/abs/1410.8849}{\texttt{arXiv:1410.8849}}.

\bibitem{Ball:2017nwa}
\hrefCMSnoop {}{{NNPDF} Collaboration, ``Parton distributions from
  high-precision collider data'',} \textit{ Eur. Phys. J. C} \textbf{ 77}
  (2017) 663,
  \href{http://dx.doi.org/10.1140/epjc/s10052-017-5199-5}{\doi{10.1140/epjc/s10052-017-5199-5}},
\href{http://www.arXiv.org/abs/1706.00428}{\texttt{arXiv:1706.00428}}.

\bibitem{Agostinelli:2002hh}
\hrefCMSnoop {}{{GEANT4} Collaboration, ``{\GEANTfour}---a simulation
  toolkit'',} \textit{ Nucl. Instrum. Meth. A} \textbf{ 506} (2003) 250,
\href{http://dx.doi.org/10.1016/S0168-9002(03)01368-8}{\doi{10.1016/S0168-9002(03)01368-8}}.

\bibitem{Melia:2011tj}
\hrefCMSnoop {}{T.~Melia, P.~Nason, R.~Rontsch, and G.~Zanderighi,
  ``{W$^+$W$^-$, WZ and ZZ production in the {POWHEG} {BOX}}'',} \textit{ JHEP}
  \textbf{ 11} (2011) 078,
  \href{http://dx.doi.org/10.1007/JHEP11(2011)078}{\doi{10.1007/JHEP11(2011)078}},
\href{http://www.arXiv.org/abs/1107.5051}{\texttt{arXiv:1107.5051}}.

\bibitem{Beneke:2011mq}
\hrefCMSnoop {}{M.~Beneke, P.~Falgari, S.~Klein, and C.~Schwinn, ``{Hadronic
  top-quark pair production with {NNLL} threshold resummation}'',} \textit{
  Nucl. Phys. B} \textbf{ 855} (2012) 695,
  \href{http://dx.doi.org/10.1016/j.nuclphysb.2011.10.021}{\doi{10.1016/j.nuclphysb.2011.10.021}},
\href{http://www.arXiv.org/abs/1109.1536}{\texttt{arXiv:1109.1536}}.

\bibitem{Cacciari:2011hy}
M.~Cacciari\hrefCMSnoop {}{ {et~al.}, ``{Top-pair production at hadron
  colliders with next-to-next-to-leading logarithmic soft-gluon
  resummation}'',} \textit{ Phys. Lett. B} \textbf{ 710} (2012) 612,
  \href{http://dx.doi.org/10.1016/j.physletb.2012.03.013}{\doi{10.1016/j.physletb.2012.03.013}},
\href{http://www.arXiv.org/abs/1111.5869}{\texttt{arXiv:1111.5869}}.

\bibitem{Baernreuther:2012ws}
\hrefCMSnoop {}{P.~B{\"{a}}rnreuther, M.~Czakon, and A.~Mitov,
  ``Percent-level-precision physics at the {Tevatron}: Next-to-next-to-leading
  order {QCD} corrections to {$\qqbar\to\ttbar +X$}'',} \textit{ Phys. Rev.
  Lett.} \textbf{ 109} (2012) 132001,
  \href{http://dx.doi.org/10.1103/PhysRevLett.109.132001}{\doi{10.1103/PhysRevLett.109.132001}},
\href{http://www.arXiv.org/abs/1204.5201}{\texttt{arXiv:1204.5201}}.

\bibitem{Czakon:2012zr}
\hrefCMSnoop {}{M.~Czakon and A.~Mitov, ``{NNLO} corrections to top-pair
  production at hadron colliders: the all-fermionic scattering channels'',}
  \textit{ JHEP} \textbf{ 12} (2012) 054,
  \href{http://dx.doi.org/10.1007/JHEP12(2012)054}{\doi{10.1007/JHEP12(2012)054}},
\href{http://www.arXiv.org/abs/1207.0236}{\texttt{arXiv:1207.0236}}.

\bibitem{Czakon:2012pz}
\hrefCMSnoop {}{M.~Czakon and A.~Mitov, ``{NNLO} corrections to top pair
  production at hadron colliders: the quark-gluon reaction'',} \textit{ JHEP}
  \textbf{ 01} (2013) 080,
  \href{http://dx.doi.org/10.1007/JHEP01(2013)080}{\doi{10.1007/JHEP01(2013)080}},
\href{http://www.arXiv.org/abs/1210.6832}{\texttt{arXiv:1210.6832}}.

\bibitem{Czakon:2013goa}
\hrefCMSnoop {}{M.~Czakon, P.~Fiedler, and A.~Mitov, ``Total top-quark
  pair-production cross section at hadron colliders through {$O(\alpS^4)$}'',}
  \textit{ Phys. Rev. Lett.} \textbf{ 110} (2013) 252004,
  \href{http://dx.doi.org/10.1103/PhysRevLett.110.252004}{\doi{10.1103/PhysRevLett.110.252004}},
\href{http://www.arXiv.org/abs/1303.6254}{\texttt{arXiv:1303.6254}}.

\bibitem{Gavin:2012sy}
\hrefCMSnoop {}{R.~Gavin, Y.~Li, F.~Petriello, and S.~Quackenbush, ``{W}
  physics at the {LHC} with {FEWZ} 2.1'',} \textit{ Comput. Phys. Commun.}
  \textbf{ 184} (2013) 208,
  \href{http://dx.doi.org/10.1016/j.cpc.2012.09.005}{\doi{10.1016/j.cpc.2012.09.005}},
\href{http://www.arXiv.org/abs/1201.5896}{\texttt{arXiv:1201.5896}}.

\bibitem{Gavin:2010az}
\hrefCMSnoop {}{R.~Gavin, Y.~Li, F.~Petriello, and S.~Quackenbush, ``{FEWZ}
  2.0: A code for hadronic {Z} production at next-to-next-to-leading order'',}
  \textit{ Comput. Phys. Commun.} \textbf{ 182} (2011) 2388,
  \href{http://dx.doi.org/10.1016/j.cpc.2011.06.008}{\doi{10.1016/j.cpc.2011.06.008}},
\href{http://www.arXiv.org/abs/1011.3540}{\texttt{arXiv:1011.3540}}.

\bibitem{Lindert:2017olm}
\hrefCMSnoop {}{J.~M. Lindert {et~al.}, ``Precise predictions for {V}+jets dark
  matter backgrounds'',} \textit{ Eur. Phys. J. C} \textbf{ 77} (2017) 829,
  \href{http://dx.doi.org/10.1140/epjc/s10052-017-5389-1}{\doi{10.1140/epjc/s10052-017-5389-1}},
\href{http://www.arXiv.org/abs/1705.04664}{\texttt{arXiv:1705.04664}}.

\bibitem{Bolognesi:2012mm}
S.~Bolognesi\hrefCMSnoop {}{ {et~al.}, ``Spin and parity of a single-produced
  resonance at the {LHC}'',} \textit{ Phys. Rev. D} \textbf{ 86} (2012) 095031,
  \href{http://dx.doi.org/10.1103/PhysRevD.86.095031}{\doi{10.1103/PhysRevD.86.095031}},
  \href{http://www.arXiv.org/abs/1208.4018}{\texttt{arXiv:1208.4018}}.

\bibitem{Oliveira:2014kla}
\hrefCMSnoop {}{A.~Oliveira, ``Gravity particles from warped extra dimensions,
  predictions for {LHC}'',} 2014.
\href{http://www.arXiv.org/abs/1404.0102}{\texttt{arXiv:1404.0102}}.

\bibitem{CMS-PRF-14-001}
\hrefCMSnoop {}{{CMS Collaboration}, ``Particle-flow reconstruction and global
  event description with the {CMS} detector'',} \textit{ JINST} \textbf{ 12}
  (2017) P10003,
  \href{http://dx.doi.org/10.1088/1748-0221/12/10/P10003}{\doi{10.1088/1748-0221/12/10/P10003}},
\href{http://www.arXiv.org/abs/1706.04965}{\texttt{arXiv:1706.04965}}.

\bibitem{Cacciari:2008gp}
\hrefCMSnoop {}{M.~Cacciari, G.~P. Salam, and G.~Soyez, ``The anti-\kt jet
  clustering algorithm'',} \textit{ JHEP} \textbf{ 04} (2008) 063,
  \href{http://dx.doi.org/10.1088/1126-6708/2008/04/063}{\doi{10.1088/1126-6708/2008/04/063}},
\href{http://www.arXiv.org/abs/0802.1189}{\texttt{arXiv:0802.1189}}.

\bibitem{Cacciari:2011ma}
\hrefCMSnoop {}{M.~Cacciari, G.~P. Salam, and G.~Soyez, ``{FastJet} user
  manual'',} \textit{ Eur. Phys. J. C} \textbf{ 72} (2012) 1896,
  \href{http://dx.doi.org/10.1140/epjc/s10052-012-1896-2}{\doi{10.1140/epjc/s10052-012-1896-2}},
\href{http://www.arXiv.org/abs/1111.6097}{\texttt{arXiv:1111.6097}}.

\bibitem{Khachatryan:2015hwa}
\hrefCMSnoop {}{{CMS Collaboration}, ``{Performance of electron reconstruction
  and selection with the {CMS} detector in proton-proton collisions at
  \ensuremath{\sqrt{s} = 8\TeV}}'',} \textit{ JINST} \textbf{ 10} (2015)
  P06005,
  \href{http://dx.doi.org/10.1088/1748-0221/10/06/P06005}{\doi{10.1088/1748-0221/10/06/P06005}},
  \href{http://www.arXiv.org/abs/1502.02701}{\texttt{arXiv:1502.02701}}.

\bibitem{Sirunyan:2018fpa}
\hrefCMSnoop {}{{CMS Collaboration}, ``Performance of the {CMS} muon detector
  and muon reconstruction with proton-proton collisions at {$\sqrt{s} =
  13$\TeV}'',} \textit{ JINST} \textbf{ 13} (2018) P06015,
  \href{http://dx.doi.org/10.1088/1748-0221/13/06/P06015}{\doi{10.1088/1748-0221/13/06/P06015}},
\href{http://www.arXiv.org/abs/1804.04528}{\texttt{arXiv:1804.04528}}.

\bibitem{Rehermann:2010vq}
\hrefCMSnoop {}{K.~Rehermann and B.~Tweedie, ``Efficient identification of
  boosted semileptonic top quarks at the {LHC}'',} \textit{ JHEP} \textbf{ 03}
  (2011) 059,
  \href{http://dx.doi.org/10.1007/JHEP03(2011)059}{\doi{10.1007/JHEP03(2011)059}},
\href{http://www.arXiv.org/abs/1007.2221}{\texttt{arXiv:1007.2221}}.

\bibitem{Khachatryan:2015iwa}
\hrefCMSnoop {}{{CMS Collaboration}, ``Performance of photon reconstruction and
  identification with the {CMS} detector in proton-proton collisions at
  {$\sqrt{s} = 8\TeV$}'',} \textit{ JINST} \textbf{ 10} (2015) P08010,
  \href{http://dx.doi.org/10.1088/1748-0221/10/08/P08010}{\doi{10.1088/1748-0221/10/08/P08010}},
\href{http://www.arXiv.org/abs/1502.02702}{\texttt{arXiv:1502.02702}}.

\bibitem{Khachatryan:2016kdb}
\hrefCMSnoop {}{{CMS Collaboration}, ``Jet energy scale and resolution in the
  {CMS} experiment in pp collisions at 8 {TeV}'',} \textit{ JINST} \textbf{ 12}
  (2017) P02014,
  \href{http://dx.doi.org/10.1088/1748-0221/12/02/P02014}{\doi{10.1088/1748-0221/12/02/P02014}},
\href{http://www.arXiv.org/abs/1607.03663}{\texttt{arXiv:1607.03663}}.

\bibitem{Bertolini:2014bba}
\hrefCMSnoop {}{D.~Bertolini, P.~Harris, M.~Low, and N.~Tran, ``{Pileup per
  particle identification}'',} \textit{ JHEP} \textbf{ 10} (2014) 059,
  \href{http://dx.doi.org/10.1007/JHEP10(2014)059}{\doi{10.1007/JHEP10(2014)059}},
\href{http://www.arXiv.org/abs/1407.6013}{\texttt{arXiv:1407.6013}}.

\bibitem{Sirunyan:2020foa}
\hrefCMSnoop {}{{CMS Collaboration}, ``{Pileup mitigation at CMS in 13 TeV
  data}'',} \textit{ JINST} \textbf{ 15} (2020) P09018,
  \href{http://dx.doi.org/10.1088/1748-0221/15/09/p09018}{\doi{10.1088/1748-0221/15/09/p09018}},
  \href{http://www.arXiv.org/abs/2003.00503}{\texttt{arXiv:2003.00503}}.

\bibitem{cms-pas-jme-10-003}
\href {http://cdsweb.cern.ch/record/1279362}{{CMS Collaboration}, ``Jet
  performance in pp collisions at $\sqrt{s}=7$~{TeV}'',} CMS Physics Analysis
  Summary CMS-PAS-JME-10-003, 2010.

\bibitem{CMS-PAS-JME-16-003}
\href {http://cds.cern.ch/record/2256875}{{{CMS}} Collaboration, ``Jet
  algorithms performance in 13 {TeV} data'',} CMS Physics Analysis Summary
  CMS-PAS-JME-16-003, 2017.

\bibitem{Dasgupta:2013ihk}
\hrefCMSnoop {}{M.~Dasgupta, A.~Fregoso, S.~Marzani, and G.~P. Salam,
  ``{Towards an understanding of jet substructure}'',} \textit{ JHEP} \textbf{
  09} (2013) 029,
  \href{http://dx.doi.org/10.1007/JHEP09(2013)029}{\doi{10.1007/JHEP09(2013)029}},
  \href{http://www.arXiv.org/abs/1307.0007}{\texttt{arXiv:1307.0007}}.

\bibitem{Larkoski:2014wba}
\hrefCMSnoop {}{A.~J. Larkoski, S.~Marzani, G.~Soyez, and J.~Thaler, ``{Soft
  drop}'',} \textit{ JHEP} \textbf{ 05} (2014) 146,
  \href{http://dx.doi.org/10.1007/JHEP05(2014)146}{\doi{10.1007/JHEP05(2014)146}},
  \href{http://www.arXiv.org/abs/1402.2657}{\texttt{arXiv:1402.2657}}.

\bibitem{Sirunyan_2020}
\hrefCMSnoop {}{{CMS Collaboration}, ``Identification of heavy, energetic,
  hadronically decaying particles using machine-learning techniques'',}
  \textit{ JINST} \textbf{ 15} (Jun, 2020) P06005,
  \href{http://dx.doi.org/10.1088/1748-0221/15/06/p06005}{\doi{10.1088/1748-0221/15/06/p06005}},
  \href{http://www.arXiv.org/abs/2004.08262}{\texttt{arXiv:2004.08262}}.

\bibitem{Thaler:2010tr}
\hrefCMSnoop {}{J.~Thaler and K.~Van~Tilburg, ``Identifying boosted objects
  with {N-subjettiness}'',} \textit{ JHEP} \textbf{ 03} (2011) 015,
  \href{http://dx.doi.org/10.1007/JHEP03(2011)015}{\doi{10.1007/JHEP03(2011)015}},
  \href{http://www.arXiv.org/abs/1011.2268}{\texttt{arXiv:1011.2268}}.

\bibitem{Sirunyan:2017ezt}
\hrefCMSnoop {}{{CMS Collaboration}, ``Identification of heavy-flavour jets
  with the {CMS} detector in pp collisions at 13~{TeV}'',} \textit{ JINST}
  \textbf{ 13} (2018) P05011,
  \href{http://dx.doi.org/10.1088/1748-0221/13/05/P05011}{\doi{10.1088/1748-0221/13/05/P05011}},
\href{http://www.arXiv.org/abs/1712.07158}{\texttt{arXiv:1712.07158}}.

\bibitem{Sirunyan:2019kia}
\hrefCMSnoop {}{{CMS Collaboration}, ``{Performance of missing transverse
  momentum reconstruction in proton-proton collisions at $\sqrt{s} =$ 13~{TeV}
  using the {CMS} detector}'',} \textit{ JINST} \textbf{ 14} (2019) P07004,
  \href{http://dx.doi.org/10.1088/1748-0221/14/07/P07004}{\doi{10.1088/1748-0221/14/07/P07004}},
  \href{http://www.arXiv.org/abs/1903.06078}{\texttt{arXiv:1903.06078}}.

\bibitem{Sirunyan:2021qkt}
\hrefCMSnoop {}{{CMS Collaboration}, ``{Precision luminosity measurement in
  proton-proton collisions at $\sqrt{s} =$ 13 TeV in 2015 and 2016 at CMS}'',}
  2021. \href{http://www.arXiv.org/abs/2104.01927}{\texttt{arXiv:2104.01927}}.
  Accepted by \textit{Eur. Phys. J. C}.

\bibitem{CMS-PAS-LUM-17-004}
\href {https://cds.cern.ch/record/2621960/}{{CMS Collaboration}, ``{CMS}
  luminosity measurement for the 2017 data-taking period at $\sqrt{s}$ = 13
  {TeV}'',} cms physics analysis summary, 2018.

\bibitem{CMS-PAS-LUM-18-002}
\href {https://cds.cern.ch/record/2676164/}{{CMS Collaboration}, ``{CMS}
  luminosity measurement for the 2018 data-taking period at $\sqrt{s}$ = 13
  {TeV}'',} cms physics analysis summary, 2019.

\bibitem{VISCHIA2020100046}
\hrefCMSnoop {}{P.~Vischia, ``Reporting results in high energy physics
  publications: A manifesto'',} \textit{ Rev. Phys.} \textbf{ 5} (2020) 100046,
  \href{http://dx.doi.org/10.1016/j.revip.2020.100046}{\doi{10.1016/j.revip.2020.100046}},
  \href{http://www.arXiv.org/abs/1904.11718}{\texttt{arXiv:1904.11718}}.

\bibitem{bib-cls}
\hrefCMSnoop {}{A.~L. Read, ``Presentation of search results: the {\CLs}
  technique'',} \textit{ J. Phys. G} \textbf{ 28} (2002) 2693,
\href{http://dx.doi.org/10.1088/0954-3899/28/10/313}{\doi{10.1088/0954-3899/28/10/313}}.

\bibitem{Cowan:2010js}
\hrefCMSnoop {}{G.~Cowan, K.~Cranmer, E.~Gross, and O.~Vitells, ``{Asymptotic
  formulae for likelihood-based tests of new physics}'',} \textit{ Eur. Phys.
  J. C} \textbf{ 71} (2011) 1554,
  \href{http://dx.doi.org/10.1140/epjc/s10052-011-1554-0}{\doi{10.1140/epjc/s10052-011-1554-0}},
  \href{http://www.arXiv.org/abs/1007.1727}{\texttt{arXiv:1007.1727}}.
[Erratum: \DOI{10.1140/epjc/s10052-013-2501-z}].

\end{thebibliography}\endgroup
